\documentclass[12pt]{amsart}

\usepackage[all, cmtip]{xy}

\usepackage{graphicx, subfig}
\usepackage{epsfig}
\usepackage{amscd}
\usepackage[mathscr]{eucal}
\usepackage{amssymb}
\usepackage{amsxtra}
\usepackage{amsmath}
\usepackage{latexsym}
\usepackage[all]{xy}
\usepackage{enumerate}
\usepackage{mathrsfs}
\usepackage{color}
\usepackage{rotating}
\usepackage[colorlinks=true,linkcolor=red,urlcolor=green,citecolor=blue]{hyperref}
\usepackage[hyperpageref]{backref}

\usepackage{color}
\theoremstyle{plain}

\theoremstyle{definition}

\theoremstyle{remark}

\theoremstyle{plain}

\begin{document}

\title[]{Fourth order Superintegrable systems separating in Cartesian coordinates\\I. Exotic quantum potentials}
\date{\today}

\author[]{Ian Marquette, Masoumeh Sajedi, Pavel Winternitz}

\address{Ian Marquette, School of Mathematics and Physics\\
 The University of Queensland, Brisbane, QLD 4072, Australia} \email{i.marquette@uq.edu.au}
\address{Masoumeh Sajedi, D\'epartement de math\'ematiques et de statistiques\\
 Universit\'e de Montr\'eal, C.P.6128 succ. Centre-Ville, Montr\'eal (QC) H3C 3J7, Canada} \email{sajedim@dms.umontreal.ca}
\address{Pavel Winternitz, Centre de recherches math\'ematiques and D\'epartement de math\'ematiques et de statistiques\\
 Universit\'e de Montr\'eal, C.P.6128 succ. Centre-Ville, Montr\'eal (QC) H3C 3J7, Canada} \email{wintern@crm.umontreal.ca}
  
%

 \begin{abstract}
A study is presented of two-dimensional superintegrable systems separating in Cartesian coordinates and allowing an integral of motion that is a fourth order polynomial in the momenta. All quantum mechanical potentials that do not satisfy any linear differential equation are found. They do however satisfy nonlinear ODEs. We show that these equations always have the Painlev\'e property and integrate them in terms of known Painlev\'e transcendents or elliptic functions.\\

 \end{abstract}

\maketitle


\section{Introduction}

This article is part of a general program the aim of which is to derive, classify, and solve the equations of motion of superintegrable systems with integrals of motion that are polynomials of finite order N in the components of linear momentum. So far, we are concentrating on superintegrable systems with Hamiltonians of the form
\begin{equation}\label{H}
H=\frac{1}{2}(p_1^2+p_2^2)+V(x,y),
\end{equation} 
in two dimensional Euclidean space $E_2$. In classical mechanics, $p_1$ and $p_2$ are the canonical momenta conjugate to the Cartesian coordinates $x$ and $y$. In quantum mechanics, we have
\begin{equation}\label{p,L}
p_1=-i\hbar \partial_x, \quad p_2=-i\hbar \partial_y, \quad L_3=xp_2-yp_1.
\end{equation}
The angular momentum $L_3$ is introduced because it will be needed below.\\
We recall that a superintegrable system has more integrals of motion than degrees of freedom (see [MPW13] for a recent review with an extensive list of references). More precisely, a classical Hamiltonian system with n degrees of freedom is integrable if it allows n integrals of motion $\{X_1,X_2,...X_n\}$ (including the Hamiltonian) that are in involution, are well defined functions on the phase space and are functionally independent. It is superintegrable if further functionally independent integrals exist, $\{Y_1, Y_2, ...,Y_k\}$ with $1 \leq k \leq n-1.$ The value $k=1$ corresponds to "minimal superintegrability," $k=n-1$ to "maximal superintegrability." In quantum mechanics, the integrals are operators in the enveloping algebra of the Heisenberg algebra (or in some generalization of the enveloping algebra). In this article we assume that all integrals are polynomials in the momenta of the order $1 \leq j \leq N,$ and at least one of them is of order $N.$ We require the integrals to be algebraically independent, i.e no Jordan polynomial (completely symmetric) formed out of the $n+k$ integrals of motion can vanish identically.\\
In classical mechanics, all bounded trajectories in a maximally superintegrable system are closed \cite{Nekhoroshev:actangl}, and the motion is periodic. In quantum mechanics, it has been conjectured by Tempesta, Turbiner and Winternitz  \cite{TWT:exact} that all maximally superintegrable systems are exactly solvable. This means that the bound states spectra can  be calculated algebraically and their wave functions expressed as polynomials in some appropriate variables (multiplied by an overall gauge factor).\\ 
The best known superintegrable systems in $E_n,\; n \geq 2,$ correspond to the Kepler-Coulomb potential $V=\frac{\alpha}{r}$ (see \cite{Fock:Fo, Bargmann:Ba}) and the isotropic harmonic oscillator $V=\alpha r^2$ (see \cite{Hill:deg, Moshinsky:HO}).\\
A sizable recent literature on superintegrable systems has been published. It includes theoretical studies of such systems in
Riemannian and pseudo-Riemannian spaces of arbitrary dimensions and with integrals of arbitrary order. The potentials are either scalar ones, or
may involve vector potentials, or particles with spin \cite{Carinena:17, Neg:Polyhei, Desilets:12, Gungor:14, Hakobyan:15, Nikitin:15, Nikitin:14, Ranada:15, Snobl:15, Tremblay:09}. For recent applications of superintegrable systems in such diverse fields as particle physics, general relativity, statistical physics and the theory of orthogonal polynomials see \cite{Vincent:BI, Evnin:16, Fagotti:14, Vincent:16, Ian:16, Kurochkin:16, Mohammadi:17, Pogosyan:17}.\\
According to Bertrand's theorem, (see \cite{Bertrand:Thm, Goldstein:CM}), the only spherically symmetric potentials (in $E_3$) for which all bounded trajectories are closed are precisely $\frac{1}{r}$ and $\omega^2r^2$. Hence when searching for further superintegrable systems, we must go beyond spherically symmetrical potentials.\\
A systematic search for second order superintegrable systems in $E_2$ was started by Fri{\v{s}}, Mandrosov, Smorodinsky, Uhl{\'{\i}}{\v{r}} and Winternitz \cite{Fris:Fri} and in $E_3$ by Makarov, Smorodinsky, Valiev and Winternitz \cite{Makanov:Nonrel} , and Evans \cite{Evans:CM, Evans:SW}. A relation between second order superintegrability and multiseparability of the Schr\"{o}dinger or Hamilton-Jacobi equation was also established in these articles.\\
Most of the subsequent work was devoted to second order superintegrability (X and Y polynomials of order 2 in the momenta) and is reviewed in an article by Miller, Post and Winternitz \cite{Miller:review}. The study of third order superintegrability ($X$ of order 1 or 2, $Y$ of order 3) started in 2002 by Gravel and Winternitz \cite{Gravel:GP, Gravel:G3}, and new features were discovered. Third order integrals in classical mechanics in a complex plane were studied earlier by Drach and he found 10 such integrable systems \cite{Drach:Dr}.
The Drach systems were more recently studied by Ra\~{n}ada \cite{Ranada:Drach} and Tsiganov \cite{Tsiganov:Drach} who showed that 7 of the 10 systems are actually reducible. These 7 systems are second order superintegrable and the third order integral is a commutator (or Poisson commutator) of two second order ones.\\
The determining equations for the existence of an $N$th order integral of motion in two-dimensional Euclidean space were derived by Post and Winternitz in \cite{Post:Nth}. The Planck constant $\hbar$ enters explicitly in the quantum case. The classical determining equations are obtained in the limit $\hbar \to 0.$ The classical and quantum cases differ for $N \geq 3$ and in the classical case the determining equations are much simpler. The determining equations constitute a system of partial differential equations (PDE) for the potential $V(x,y)$ and for the functions $f_{ab}(x,y)$ multiplying the monomials $p_1^ap_2^b$ in the integral of motion. If $V(x,y)$ is given, the PDEs for $f_{ab}(x,y)$ are linear. If we are searching for potentials that allow an integral of order $N$ the set of PDEs is nonlinear. A linear compatibility condition for the potential $V(x,y)$ alone was derived in \cite{Post:Nth}. It is an $N$th order PDE with polynomial coefficients also of order up to $N.$\\
An interesting phenomenon was observed when studying third order superintegrable quantum systems in $E_2$. Namely, when the potential allows a third order integral and in addition a second order one (that leads to separation of variables in either Cartesian or polar coordinates) "exotic potentials" arise, (see \cite{Gravel:GP, Gravel:G3, Tremblay:3polar}). These are potentials that do not satisfy any linear differential equation but instead satisfy nonlinear ordinary differential equations (ODEs). It turned out that all the ODEs obtained in the quantum case have the Painlev\'e property. That means that the general solution of these equations has no movable critical singularities (see \cite{Ince:ode, Painleve:Pain, Gambier:Pain, Conte:R99, Conte:B08}). It can hence be expanded into a Laurent series with a finite number of negative powers. The separable potentials were then expressed in terms of elliptic functions, or known (second order) Painlev\'e transcendents (i.e. the solutions of the Painlev\'e equations  \cite[page 345]{Ince:ode}).\\
We conjecture that this is a general feature of quantum superintegrable systems in two-dimensional Euclidean spaces. Namely, that if they allow an integral of motion of order $N\geq 3$ and also allow the separation of variables in Cartesian or polar coordinates, they will involve potentials that are solutions of ordinary differential equations that have the Painlev\'e property. All linear equations have this property by default, they have no movable singularities at all. Exotic potentials, on the other hand, are solutions of a genuinely nonlinear ODEs that have the Painlev\'e property.\\
The specific aim of this article is to test the above conjecture for superintegrable systems allowing one fourth order integral of motion and one second order one that leads to the separation of variables in Cartesian coordinates. We will determine all such exotic potentials and obtain their explicit expressions.\\
In Section $2$, we present the set of $6$ determining equations for the fourth order integral $Y_L$ as well as a linear compatibility condition for $4$ of these equations. This is a fourth order linear PDE for the potential $V(x,y)$. In Section $3$, we impose the existence of an additional second order "Cartesian" integral that restricts the form of the potential to $V(x,y)=V_1(x)+V_2(y)$. The linear compatibility condition then reduces to two linear ODEs for $V_1(x)$ and two for $V_2(y)$. Section $4$ is an auxiliary one. In it we review same basic facts about nonlinear equations with the Painlev\'e property that will be needed below (they come mainly from the references \cite{Bureau:39, Bureau:64I, Bureau:64II, Bureau:SD, Chazy:CI, Chalkley:fuchs, Cosgrove:chazy, Cosgrove:SD, Fuchs:1st}). The main original results of this paper are contained in Section $5$. We impose that the linear equation for at least one of the functions $V_1(x)$ or $V_2(y)$ be satisfied trivially (otherwise the potential would not be exotic.) This greatly simplifies the form of the integral $Y$ (6 out of 10 free constants must vanish). The remaining linear and nonlinear determining equations can be solved exactly and completely. As expected, we find that the potentials satisfy nonlinear equations that pass the Painlev\'e test introduced by Ablowitz, Ramani, and Segur \cite{Ablowitz:Pain} (see also Kowalevski \cite{Kowalevski:PT89} and Gambier \cite{Gambier:Pain}). Using the results of \cite{Chazy:CI, Bureau:SD, Cosgrove:SD, Cosgrove:chazy}, we integrate these 4th order ODEs in terms of the original 6 Painlev\'e transcendents, elliptic functions, or solutions of linear equations. In Section $6$, we study the classical analogs of exotic potentials. They satisfy first order ODEs that are polynomials of second degree in the derivative. Section $7$ is devoted to conclusions and future outlook.


\section{DETERMINING EQUATIONS AND LINEAR COMPATIBILITY CONDITION FOR A FOURTH ORDER INTEGRAL}

The determining equations for fourth-order classical and quantum integrals of motion were derived earlier by Post and Winternitz \cite{Post:fourth} and they are a special case of $N$th order ones given in \cite{Post:Nth}. In the quantum case, the integral is $Y^{(4)}=Y:$
\begin{align}\label{Y}
Y=\sum_{j+k+l=4} \frac{A_{jkl}}{2} \{L_3^j,p_1^k p_2^l\}+\frac{1}{2}(\{g_1(x,y),p_1^2\}+\{g_2(x,y),p_1p_2\}+\{g_3(x,y),p_2^2\})+l(x,y),\nonumber\\
\end{align}
where $A_{jkl}$ are real constants, the brackets $\{.,.\}$ denote anti-commutators and the Hermitian operators $p_1,p_2$ and $L_3$ are given in (\ref{p,L}). The functions $g_1(x,y), g_2(x,y), g_3(x,y),$ and $l(x,y)$ are real and the operator $Y$ is self adjoint. Equation (\ref{Y}) is also valid in classical mechanics where $p_1, p_2$ are the canonical momenta conjugate to $x$ and $y$, respectively (and the symmetrization becomes irrelevant).\\
The commutation relation $[H,Y]=0$ with $H$ in (\ref{H}) provides the determining equations
\begin{subequations}\label{det1}
\begin{align}
g_{1,x}=4f_1V_x+f_2V_y\label{det1a}\\
g_{2,x}+g_{1,y}=3f_2V_x+2f_3V_y\label{det1b}\\
g_{3,x}+g_{2,y}=2f_3V_x+3f_4V_y\label{det1c}\\
g_{3,y}=f_4V_x+4f_5V_y,\label{det1d}
\end{align}
\end{subequations}

and 
\begin{subequations}\label{l1}
\begin{align}
\ell_{x}=&2g_1V_x+g_{2}V_y+\frac{\hbar^2}{4}\bigg((f_2+f_4)V_{xxy}-4(f_1-f_5)V_{xyy}-(f_2+f_4)V_{yyy}\nonumber\\
&+(3f_{2,y}-f_{5,x})V_{xx}-(13f_{1,y}+f_{4,x})V_{xy}-4(f_{2,y}-f_{5,x})V_{yy}\nonumber\\
& -2(6A_{400}x^2+62A_{400}y^2+3A_{301}x-29A_{310}y+9A_{220}+3A_{202})V_x\nonumber\\
& +2(56A_{400}xy-13A_{310}x+13A_{301}y-3A_{211})V_y\bigg),\label{lx}\\
\ell_{y}=&g_{2}V_x+2g_{3}V_y+\frac{\hbar^2}{4}\bigg(-(f_2+f_4)V_{xxx}+4(f_1-f_5)V_{xxy}+(f_2+f_4)V_{xyy}\nonumber\\
&+4(f_{1,y}-f_{4,x})V_{xx}-(f_{2,y}+13f_{5,x})V_{xy}-(f_{1,y}-3f_{4,x})V_{yy}\nonumber\\
&+2(56A_{400}xy-13A_{310}x+13A_{301}y-3A_{211})V_x\nonumber\\
&-2(62A_{400}x^2+6A_{400}y^2+29A_{301}x-3A_{310}y+9A_{202}+3A_{220})V_y\bigg).\label{ly}
\end{align}
\end{subequations}
The quantities $f_i,\; i=1,2,..,5$ are polynomials, obtained from the highest order term in the condition $[H,Y]=0$, and explicitly we have
\begin{align}\label{fi}
&f_1=A_{400}y^4-A_{310}y^3+A_{220}y^2-A_{130}y+A_{040}\nonumber\\
&f_2=-4A_{400}xy^3-A_{301}y^3+3A_{310}xy^2+A_{211}y^2-2A_{220}xy-A_{121}y+A_{130}x+A_{031}\nonumber\\
&f_3=6A_{400}x^2y^2+3A_{301}xy^2-3A_{310}x^2y+A_{202}y^2-2A_{211}xy+A_{220}x^2-A_{112}y+A_{121}x+A_{022}\nonumber\\
&f_4=-4A_{400}yx^3+A_{310}x^3-3A_{301}x^2y+A_{211}x^2-2A_{202}xy+A_{112}x-A_{103}y+A_{013}\nonumber\\
&f_5=A_{400}x^4+A_{301}x^3+A_{202}x^2+A_{103}x+A_{004}.\nonumber\\
\end{align}
For a known potential the determining equations (\ref{det1}) and (\ref{l1}) form a set of 6 linear PDEs for the functions $g_1,g_2,g_3,$ and $l$. If $V$ is not known, we have a system of 6 nonlinear PDEs for $g_i,l$ and $V$. In any case the four equations (\ref{det1}) are a priori incompatible. The compatibility equation is a fourth-order linear PDE for the potential $V(x,y)$ alone, namely
\begin{equation}\label{V}
\partial_{yyy}(4f_1V_x+f_2V_y)-\partial_{xyy}(3f_2V_x+2f_3V_y)+\partial_{xxy}(2f_3V_x+3f_4V_y)-\partial_{xxx}(f_4V_x+4f_5V_y)=0.
\end{equation}
This is a special case of the $N$th order linear compatibility equation obtained in \cite{Post:Nth}. We see that the equation (\ref{V}) does not contain the Planck constant and is hence the same in quantum and classical mechanics (this is true for any $N$ \cite{Post:Nth}).
The difference between classical and quantum mechanics manifests itself in the two equations (\ref{l1}). They greatly simplify in the classical limit $\hbar \to 0$. Further compatibility conditions on the potential $V(x,y)$ can be derived for the systems (\ref{det1}) and (\ref{l1}), they will however be nonlinear. We will not go further into the problem of the fourth order integrability of the Hamiltonian (\ref{H}). Instead, we turn to the problem of superintegrability formulated in the Introduction.
\section{POTENTIALS SEPARABLE IN CARTESIAN COORDINATES}\label{sepsection}

We shall now assume that the potential in the Hamiltonian (\ref{H}) has the form
\begin{equation}\label{Vs}
V(x,y)=V_1(x)+V_2(y).
\end{equation}
This is equivalent to saying that a second order integral exists which can be taken in the form
\begin{equation}\label{X2}
X=\frac{1}{2}(p_1^2-p_2^2)+V_1(x)-V_2(y).
\end{equation}
Equivalently, we have two one dimensional Hamiltonians
\begin{equation}\label{H1,H2}
H_1=\dfrac{p_1^2}{2}+V_1(x), \quad H_2=\dfrac{p_2^2}{2}+V_2(y).
\end{equation}
We are looking for a third integral of the form (\ref{Y}) satisfying the determining equations  (\ref{det1}) and (\ref{l1}). This means that we wish to find all potentials of the form (\ref{Vs}) that satisfy the linear compatibility condition (\ref{V}). Once (\ref{Vs}) is substituted, (\ref{V}) is no longer a PDE and will split into a set of ODEs which we will solve for $V_1(x)$ and $V_2(y)$.\\
The task thus is to determine and classify all potentials of the considered form that allow the existence of at least one fourth order integral of motion. As in every classification we must avoid triviality and redundancy. Since $H_1$ and $H_2$ of (\ref{H1,H2}) are integrals, we immediately obtain 3 "trivial" fourth order integrals, namely $H_1^2, H_2^2,$ and $H_1H_2.$ The fourth order integral $Y$ of equation (\ref{Y}) can be simplified by taking linear combination with polynomials in the second order integrals $H_1$ and $H_2$ of (\ref{H1,H2}):
\begin{equation}\label{trivial}
Y \to Y'=Y+a_1H_1^2+a_2H_2^2+a_3H_1H_2+b_1H_1+b_2H_2+b_0, \quad a_i, b_i \in \mathbb{R}.
\end{equation}
Using the constants $a_1,a_2$ and $a_3$ we set
\begin{align}
A_{004}= A_{040}= A_{022} = 0,
\end{align}
in the integral $Y$ we are searching for. At a later stage we will use the constants $b_0,b_1$ and $b_2$ to eliminate certain terms in $g_1, g_3$ and $l.$\\
Other trivial fourth order integrals are more difficult to identify. They arise whenever the potential (\ref{Vs}) is lower order superintegrable i.e. in addition to (\ref{X2}), allows another second or third order integral. In such a case, the fourth order integral may be a commutator  (or Poisson commutator) of two lower order ones. Such cases must be weeded out a posteriori. In our case this is actually quite simple. The exotic potentials separating in Cartesian coordinates and allowing an additional third order integral are listed as $Q16-Q20$ in \cite{Gravel:G3}. For $4$ of them the leading terms in the third order integral $Y^{(3)}$ has the form $ap_1^3+bp_2^3$ or $ap_1^3+bp_1^2p_2$. Hence commuting $Y^{(3)}$ with a second order integral $H_1$ can not give rise to a fourth order integral.\\
The remaining case is $Q18$ with
\begin{align}\label{Vq18}
V(x,y)=a(y^2+x^2)-2 \sqrt[4]{\frac{ a^3}{2} \hbar^2}x P_4(-\sqrt[4]{\frac{2a}{\hbar^2}}x) +\sqrt{\frac{a}{2}} \hbar( \epsilon  P_4'(-\sqrt[4]{\frac{2a}{\hbar^2}}x)+ P_4^2(-\sqrt[4]{\frac{2a}{\hbar^2}}x)), \quad \epsilon=\pm 1,
\end{align}
and integral
$$Y^{(3)}=\{L_3,p_1^2\}+\{ax^2y-3yV_1,p_1\}-\frac{1}{2a}\{\frac{\hbar ^2}{4}V_{1xxx}+(ax^2-3V_1)V_{1x},p_2\}.$$
Commuting $Y^{(3)}$ with $H_1$ we obtain a fourth order integral
\begin{align}\label{intq18}
Y^{(4)}=2p_1^3p_2+...
\end{align}
Hence the potential (\ref{Vq18}) must appear (and does appear) in our present study, but the existence of (\ref{intq18}) is a "trivial" consequence of third order superintegrability. However, an integral of the type (\ref{intq18}) may appear for more general potentials than (\ref{Vq18}).\\
Two potentials will be considered equivalent if and only if they differ at most by translations of $x$ and $y$.\\
Substituting (\ref{Vs}) into the compatibility condition (\ref{V}), we obtain a linear condition, relating the functions $V_1(x)$ and $V_2(y)$
\begin{align}\label{V1+V2}
&(-60 A_{310}+240yA_{400})V_1'(x)+(-20A_{211}+60y A_{301}-60xA_{310}+240xyA_{400})V_1''(x)+\nonumber\\
&(-5A_{112}+10yA_{202}-10xA_{211}+30xyA_{301}-15x^2A_{310}+60x^2y A_{400})V_1^{(3)}(x)+\nonumber\\
&(-A_{013}+yA_{103}-xA_{112}+2xyA_{202}-x^2A_{211}+3x^2yA_{301}-x^3A_{310}+4x^3yA_{400})V_1^{(4)}(x)+\nonumber\\
&(-60A_{301}-2140xA_{400})V_2'(y)+(20A_{211}-60yA_{301}+60xA_{310}-240xyA_{400})V_2''(y)+\nonumber\\
&(-5A_{121}++10yA_{211}-10xA_{220}-15y^2A_{301}+30xyA_{310}-60xy^2A_{400})V_2^{(3)}(y)+\nonumber\\
&(A_{031}-yA_{121}+xA_{130}+y^2 A_{211}-2xyA_{220}-y^3A_{301}+3xy^2A_{310}-4xy^3A_{400})V_2^{(4)}(y)=0.\nonumber\\
\end{align}
It should be stressed that this is no longer a PDE, since the unknown functions $V_1(x)$ and $V_2(y)$ both depend on one variable only.\\
We differentiate (\ref{V1+V2}) twice with respect to $x$ and thus eliminate $V_2(y)$ from the equation. The resulting equation for $V_1(x)$ splits into two linear ODEs (since the coefficients contain terms proportional to $y^0,$ and $y^1$), namely

\begin{subequations}\label{*(1&2)}
\begin{align}
&210A_{310}V_1^{(3)}(x)+42(A_{211}+3A_{310}x)V_1^{(4)}(x)+7(A_{112}+2A_{211}x+3A_{310}x^2)V_1^{(5)}(x)\nonumber\\
&+(A_{013}+A_{112}x+A_{211}x^2+A_{310}x^3)V_1^{(6)}(x)=0,\nonumber\\\label{*(1&2)a}\\
&840A_{400}V_1^{(3)}(x)+(126A_{301}+504A_{400}x)V_1^{(4)}(x)+14(A_{202}+3A_{301}x+6A_{400}x^2)V_1^{(5)}(x)\nonumber\\
&+(A_{103}+2A_{202}x+3A_{301}x^2+4A_{400}x^3)V_1^{(6)}(x)=0.\label{*(1&2)b}
\end{align}
\end{subequations}
Similarly, differentiating (\ref{V1+V2}) with respect to $y$ we obtain two linear ODEs for $V_2(y),$

\begin{subequations}\label{**(1&2)}
\begin{align}
&210A_{301}V_2^{(3)}(y)-42(A_{211}-3A_{301}y)V_2^{(4)}(y)+7(A_{121}-2A_{211}y+3A_{301}y^2)V_2^{(5)}(y)\nonumber\\
&-(A_{031}-A_{121}y+A_{211}y^2-A_{301}y^3)V_2^{(6)}(y)=0,\nonumber\\\label{**(1&2)a}\\
&840A_{400}V_2^{(3)}(y)-(126A_{310}-504A_{400}y)V_2^{(4)}(y)+14(A_{220}-3A_{310}y+6A_{400}y^2)V_2^{(5)}(y)\nonumber\\
&-(A_{130}-2A_{220}y+3A_{310}y^2-4A_{400}y^3)V_2^{(6)}(y)=0.\label{**(1&2)b}
\end{align}
\end{subequations}
The compatibility condition $\ell_{xy}=\ell_{yx}$, for (\ref{lx}) and (\ref{ly}) implies
\begin{align}\label{lxy-lyx}
&-g_2 V_1''(x)+g_2 V_2''(y)+(2g_{1y}-g_{2x}) V_1'(x)+(g_{2y}-2g_{3x})V_2'(y)+\nonumber\\
&\frac{\hbar^2}{4}\bigg((f_2+f_4)(V_1^{(4)}-V_2^{(4)})+(f_{2x}-4f_1'(y))V_1^{(3)}+(4f_5'(x)-5f_{2y}-f_{4y})V_2^{(3)}\nonumber\\
&+(3f_{2yy}+4f_{4xx}+6A_{211}-26A_{301}y+26A_{310}x-112A_{400}xy) V_1''\nonumber\\
&-(4f_{2yy}+3f_{4xx}+6A_{211}-26A_{301}y+26A_{310}x-112A_{400}xy)V_2''\nonumber\\
&+(84A_{310}-360A_{400}y)V_1'+(84A_{310}+360A_{400}y)V_2'\bigg)=0.
\end{align}
This equation, contrary to (\ref{*(1&2)}) and (\ref{**(1&2)}), is nonlinear since it still involves the unknown functions $g_1, g_2,$ and $g_3$, (in addition to $V_1(x)$ and $V_2(y)$).\\
Our next task is to solve equations (\ref{*(1&2)}) and (\ref{**(1&2)}) and ultimately also (\ref{lxy-lyx}) and the other determining equations. The starting point is given by the linear compatibility conditions (\ref{*(1&2)}) and (\ref{**(1&2)}) for $V_1(x)$ and $V_2(y)$. These are third order linear ODEs for the functions $W_1(x)=V_1^{(3)}(x),$ and $W_2(y)=V_2^{(3)}(y).$ They have polynomial coefficients and are easy to solve. Once the potentials are known, the whole problem becomes linear. However, the coefficients $A_{jkl}$ (in the integral (\ref{Y}) and in (\ref{*(1&2)}) and (\ref{**(1&2)})) may be such that the equations (\ref{*(1&2)}) or (\ref{**(1&2)}) vanish identically. Then the equations provide no information. This may lead to exotic potentials not satisfying any linear equation at all. In a previous study \cite{Gravel:GP, Gravel:G3} involving third order integrals, it was shown that all exotic potentials can be expressed in terms of elliptic functions or Painlev\'e transcendents. Here we will show that the same is true for integrals of order 4.\\ 
\section{ODES WITH THE PAINLEV\'E PROPERTY}

In order to study exotic potentials $V(x,y)=V_1(x)+V_2(y),$ allowing fourth order integrals of motion in quantum mechanics we must first recall some known results on Painlev\'e type equations.
\subsection{THE PAINLEV\'E PROPERTY, PAINLEV\'E TEST AND THE CLASSIFICATION OF PAINLEV\'E TYPE EQUATIONS}

An ODE has the Painlev\'e property if its general solution has no movable branch points, (i.e. branch points whose location depends on one or more constants of integration). We shall use the Painlev\'e test in the form introduced in \cite{Ablowitz:Pain}. For a review and further developments see Conte, Fordy, and Pickering \cite{Conte:93}, Conte \cite{Conte:R99}, Conte and Musette \cite{Conte:B08, Conte:PT13}, Grammaticos and Ramani \cite{GR:97}, Hone \cite{Hone:09}, Kruskal and Clarkson \cite{Clark:92}. Passing the test is a necessary condition for having the Painlev\'e property. We shall need it only for equations of the form

\begin{align}\label{noeq}
W^{(n)}=F(y,W,W',W'',...,W^{(n-1)}),
\end{align}
where $F$ is polynomial in $W,W',W'',...,W^{(n-1)}$ and rational in $y$. The general solution must have the form of a Laurent series with a finite number of negative power terms
\begin{align}\label{laurent}
W=\Sigma_{k=0}^\infty d_k (y-y_0)^{k+p}, \; d_0 \neq 0,
\end{align} 
satisfying the requirements
\begin{enumerate}
\item The constant $p$ is a negative integer. 
\item The coefficients $d_k$ satisfy a recursion relation of the form
$$P(k)d_k=\phi_k(y_0,d_0,d_1,...,d_{k-1}),$$
where $P(k)$ is a polynomial that has $n-1$ distinct nonnegative integer zeros. The values of $k_j$ for which we have $P(k_j)=0$ are called resonances and the values of $d_k$ for $k=k_j$ are free parameters. Together with the position $y_0$ at the singularity we thus have $n$ free parameters in the general solution (\ref{laurent}).
\item A compatibility condition, also called the resonance condition:
$$\phi_k(y_0,d_0,d_1,...,d_{k-1})=0,$$
must be satisfied identically in $y_0$ and in the values of $d_{k_j}$ for all $k_j; j=1,2,...,n-1.$
\end{enumerate}
This test is a generalization of the Frobenius method used to study fixed singularities of linear ODEs (for the Frobenius method see e.g. the book by Boyce and Diprima \cite{BoyDip:EDE}). Passing the Painlev\'e test is a necessary condition only. To make it sufficient one would have to prove that the series (\ref{laurent}) has a nonzero radius of convergence and that the $n$ free parameters can be used to satisfy arbitrary initial conditions. A more practical procedure that we shall adopt is the following. Once a nonlinear ODE passes the Painlev\'e test one can try to integrate it explicitly. The Riccati equation is the only first order and first degree equation which has the Painlev\'e property. A first order algebraic differential equation of degree n $\ge 1$ has the form
\begin{align}\label{1Nd}
A_0(W,y) W'^n+A_{1}(W,y)W'^{n-1}+...+A_n(W,y)=0,
\end{align}
where $A_i$ are polynomials in $W$. When all solutions of such equation are free of movable branch points, the degree of polynomials $A_i$ must satisfy $deg(A_i) \le 2i$ for $i=0,1,2,...,n$. The necessary and sufficient conditions for such equation to have the Painlev\'e property is given by the Fuchs' theorem (Theorem1.1,\cite[page 80]{Chalkley:fuchs},proof in \cite[page 304-311]{Ince:ode}). Painlev\'e type differential equations of the first order and $n$th degree have been studied in \cite{Fuchs:1st}, \cite[chapter 13]{Ince:ode}. All such equations are either reducible to linear equations or solvable in terms of elliptic functions. Painlev\'e type second order first degree equation are of the from
$$W''=F(W',W,y),$$
where $F$ is a polynomial of degree at most 2 in $W'$, with coefficients that are rational in $W$, and analytic in $y$. They were classified by Painlev\'e and Gambier, (see \cite{Ince:ode, Davis:Pain}). They can be solved in terms of solutions of linear equations, elliptic functions or in terms of the $6$ irreducible Painlev\'e transcendents $P_I, P_{II},...,P_{VI}$.\\
Bureau initiated a study of ODEs of the form
$$A(W',W,y)W''^2+B(W',W,y)W''+C(W',W,y)=0,$$
where $A, \;B$ and $C$ are polynomials in $W,$ and $W'$ with coefficients analytic in $y$, \cite{Bureau:SD}. This work was continued by Cosgrove and Scoufis \cite{Cosgrove:SD} who constructed all Painlev\'e type ODEs of the form
$$W''^2=F(W',W,y),$$
where $F$ is rational in $W',$ and $W$ and analytic in $y$. They also succeeded in integrating all of these equations in terms of known functions (including the six original Painlev\'e transcendents).\\
We will need to integrate equations of the form (\ref{noeq}) for $n=3.$ Chazy in \cite{Chazy:CI} studied the Painlev\'e type third order differential equations in the polynomial class and proved that they have the form
\begin{equation}\label{3rd3}
W'''=aWW''+bW'^2+cW^2W'+dW^4+A(y)W''+B(y)WW'+C(y)W'+D(y)W^3+E(y)W^2+F(y)W+G(y),
\end{equation}
where $a,b,c,$ and $d$ are certain rational or algebraic numbers, and the remaining coefficients are locally analytic functions of $y$.\\
Chazy and Bureau have determined all cases for the reduced equation, obtained by using the $\alpha$-test, $(y,W) \to (y_0+\alpha y,\frac{W}{\alpha})$ when $\alpha \to 0$, \cite{Chazy:CI}. Chazy classified the reduced equations into 13 classes, denoted by Chazy class I-XIII. The list of these equations is in \cite[page 181]{Cosgrove:chazy}. Each Chazy class is a conjugacy class of differential equations under transformations of the form
$$U(Y)=\lambda(y)W+\mu (y), \; Y=\phi (y).$$
\\
In Section $5$, we will encounter some fourth order differential equations, but we always succeed in integrating them to third order ones. We then transform to a Chazy-I equation. Cosgrove in \cite{Cosgrove:chazy} introduces the canonical form for Chazy-I equation as
\begin{align}\label{chazyIr}
W'''=&-\frac{f'(y)}{f(y)}W''-\frac{2}{f^2(y)}\big(3k_1y(yW'-W)^2+k_2(yW'-W)(3yW'-W)+k_3W'(3yW'-2W)\nonumber\\
&+k_4(W')^2+2k_5y(yW'-W)+k_6(2yW'-W)+2k_7W'+k_8y+k_9\big),\nonumber\\
\end{align}
where $f(y)=k_1y^3+k_2y^2+k_3y+k_4;$ Equation (\ref{chazyIr}) admits the first integral,
\begin{align}\label{chazyIrint}
(W'')^2=&-\frac{4}{f^2(y)}\big(k_1(yW'-W)^3+k_2W'(yW'-W)^2+k_3(W')^2(yW'-W)+k_4(W')^3+k_5(yW'-W)^2\nonumber\\
&+k_6W'(yW'-W)+k_7(W')^2+k_8(yW'-W)+k_9W'+k_{10}\big),\nonumber\\
\end{align}
where $k_{10}$ is the constant of integration. In \cite{Cosgrove:SD}, Cosgrove and Scoufis give a complete classification of Painlev\'e type equations of second order and second degree. There are six classes of them, denoted by SD-I, SD-II,...,SD-VI. The equation (\ref{chazyIrint}), which is introduced as SD-I equation, splits into six canonical subcases (SD-Ia, SD-Ib, SD-Ic, SD-Id, SD-Ie, and SD-If). The solution of SD-Ia is expressed in terms of the sixth Painlev\'e transcendent. Here, we do not get any equation of this form. The solutions for the SD-Ib is expressed in terms of either the third or fifth Painlev\'e transcendent. The solutions of SD-1c, SD-Id, SD-Ie, and SD-If are, respectively, expressed in terms of the Painlev\'e IV, II, I and elliptic function \cite[page 66]{Cosgrove:SD}. These equations and their solutions appear in Section 5.
\section{SEARCH FOR EXOTIC POTENTIALS IN THE QUANTUM CASE}
\subsection{General comments}
Let us first investigate the cases that may lead to "exotic potentials", that is potentials which do not satisfy any linear differential equations. That means that either (\ref{*(1&2)}) or (\ref{**(1&2)})(or both) must be satisfied trivially. The linear ODEs (\ref{*(1&2)}) are satisfied identically if we have
\begin{align}\label{*}
A_{400}=A_{310}=A_{301}=A_{211}=A_{202}=A_{112}=A_{103}=A_{013}=0.
\end{align}
The linear ODEs (\ref{**(1&2)}) are satisfied identically if we have 
\begin{align}\label{**}
A_{400}=A_{310}=A_{301}=A_{211}=A_{220}=A_{121}=A_{130}=A_{031}=0. 
\end{align}
If (\ref{*}) and (\ref{**}) both hold then the only fourth order integrals are the trivial ones $H_1^2, H_2^2$ and $H_1H_2.$ Their existence does not assure superintegrability, it is simply a consequence of second order integrability. In other words, no fourth order superintegrable systems, satisfying (\ref{*}) and (\ref{**}) simultaneously, exist. This means that at most one of the functions $V_1(x)$ or $V_2(y)$ can be "exotic". The other one will be a solution of a linear ODE. For third order integrals both $V_1(x)$ and $V_2(y)$ could be exotic \cite{Gravel:G3}.\\
\subsection{Linear equations for $V_2(y)$ satisfied trivially}
\subsubsection{General setting and the three possible forms of $V_1(x)$}

In this case, (\ref{**}) is valid and (\ref{*}) not. The leading-order term for the nontrivial fourth order integral has the form
\begin{align}\label{V2trivially}
Y_L= A_{202} \{L_3^2, p_{2}^2\}+ A_{112} \{L_{3}, p_{1} p_{2}^2\}+ A_{103} \{L_{3}, p_{2}^3\}+2 A_{013} p_1 p_2^3.
\end{align}
Let us classify the integrals (\ref{V2trivially}) under translations. The three classes are:
\begin{align}\label{classification}
I. &A_{202} \neq 0, A_{112}=A_{103}=0.\nonumber\\
II. &A_{202}=0,  A_{112}^2+A_{103}^2\neq 0, A_{013}=0,\nonumber\\
    &IIa. A_{103} \neq 0,\nonumber\\
    &IIb. A_{103}=0, A_{112} \neq 0.\nonumber\\
III. &A_{202}=A_{112}=A_{103}=0, A_{013} \neq 0.\nonumber\\
\end{align}
The functions $f_i$ in (\ref{fi}) reduce to
\begin{align}\label{f1-5}
&f_{1}=f_{2}=0,\nonumber\\
&f_{3}(y)=A_{202}y^2-A_{112}y,\nonumber\\
&f_{4}(x,y)=-2A_{202}xy+A_{112}x-A_{103}y+A_{013},\nonumber\\
&f_{5}(x)=A_{202}x^2+A_{103}x.
\end{align}
Let us now extract all possible consequences from the determining equations (\ref{det1}). Using separability (\ref{Vs}) we obtain
\begin{align}\label{gex}
g_1(x,y)=&G_1(y),\nonumber\\
g_2(x,y)=&\big(-G_1'(y)+2(A_{202}y^2-A_{112}y)V_2'(y)\big)x+G_2(y),\nonumber\\
g_{3}(x,y)=&2(A_{202}y^2-A_{112}y) V_{1}(x)+\frac{1}{2}x(-10A_{202}xy+5A_{112}x-6A_{103}y+6A_{013})V_{2}'(y)\nonumber\\
&-x^2(A_{202}y^2-A_{112}y)V_2''(y)+\frac{1}{2}x^2G_1''(y)-xG_2'(y)+G_3(y).
\end{align}
The functions that remain to be determined are $V_1(x),V_2(y),G_1(y),G_2(y),G_3(y)$, and $l(x,y)$.\\
So far we have no information on $V_2(y)$, since equations (\ref{**(1&2)}) are satisfied trivially. The potential $V_1(x)$ must satisfy (\ref{*(1&2)a}) and (\ref{*(1&2)b}).\\
Let us substitute (\ref{f1-5}) and (\ref{gex}) into (\ref{det1d}). We obtain
\begin{align}\label{g_3y}
&(4A_{202}y-2A_{112})V_1+(2A_{202}xy-A_{112}x+A_{103}y-A_{013})V_1'-(9A_{202}x^2+7A_{103}x)V_2'\nonumber\\
&-(7A_{202}x^2y-\frac{7}{2}A_{112}x^2+3A_{103}xy-3A_{013}x)V_2''-(A_{202}x^2y^2-A_{112}x^2y)V_2^{(3)}\nonumber\\
&+G_3'(y)-xG_2''(y)+\frac{1}{2}x^2G_1^{(3)}(y)=0.
\end{align}
Differentiating (\ref{g_3y}) three times with respect to $x$ and requiring that terms proportional to $y$ and independent of $y$ vanish separately, we obtain two equations for $V_1(x)$ namely
\begin{subequations}\label{V1*}
\begin{align}
5A_{112}V_{1}^{(3)}(x)+(A_{013}+A_{112}x)V_{1}^{(4)}(x)=0,\label{V1*a}\\
10 A_{202}V_{1}^{(3)}(x)+(A_{103}+2A_{202}x)V_{1}^{(4)}(x)=0.\label{V1*b}
\end{align}
\end{subequations}
(They replace equations (\ref{*(1&2)})). These two equations imply $V_1^{(3)}=V_1^{(4)}=0$ unless we have 
\begin{align}\label{comp}
A_{112}A_{103}-2 A_{202}A_{013}=0.
\end{align}
If (\ref{comp}) is not satisfied, the only solution of (\ref{V1*}) is $V_1(x)=c_0+c_1x+c_2x^2.$ We can always put $c_0=0$. If $c_2 \neq 0$ we can translate $x$ to set $c_1=0$. Thus, with no loss of generality we can in this case put 
\begin{align}\label{V_1a}
V_1^{(a)}(x)=c_1x+c_2x^2; \quad c_1c_2=0.
\end{align}
This case will be investigated separately below in the section (\ref{potentialho}).\\
Now let us assume that (\ref{comp}) is satisfied and consider the 3 cases in (\ref{classification}) separately.\\\\
I. $A_{202} \neq 0, A_{112}=A_{103}=0, Y_L=A_{202}\{L_3^2,p_2^2\}.$\\
The condition (\ref{comp}) implies $A_{013}=0$ and from (\ref{V1*}) we obtain
\begin{align}\label{Vi}
V_1^{(b)}(x)=\dfrac{c_{-2}}{x^2}+c_1 x+c_2 x^2;\; c_{-2}\neq 0.
\end{align} 
For $c_{-2}=0, \; V_1^{(b)}$ reduces to the case $V_1^{(a)}$ of (\ref{V_1a}) .\\\\
II. $A_{202}=0, A_{112}^2+A_{103}^2 \neq 0, A_{013}=0.$\\
The condition (\ref{comp}) implies $A_{112}A_{103}=0$, so we have 2 subcases\\\\
IIa. $A_{103}\neq 0, A_{112}=0, \; Y_L=A_{103}\{L_3,p_2^3\}.$\\
The solution for (\ref{V1*}) is 
$$V_1^{(c)}(x)=c_1x+c_2x^2+c_3x^3,$$
however (\ref{g_3y}) implies $c_3=0.$ So in this case $V_1^{(c)}$ is reduced to $V_1^{(a)}.$

IIb. $A_{103}=A_{013}=0, A_{112} \neq 0, Y_L=A_{112}\{L_3,p_1p_2^2\}.$\\
The potential $V_1(x)=V_1^{(b)}(x)$ and satisfies (\ref{Vi}).\\\\
III. $A_{202}=A_{112}=A_{103}=0, A_{013}\neq 0, Y_L=2A_{013}p_1p_2^3.$\\
The potential is $V_1(x)=V_1^{(a)}(x)$ of (\ref{V_1a}).\\
Let us now return to the determining equations (\ref{l1}) and their compatibility condition (\ref{lxy-lyx}). We substitute (\ref{f1-5}) and (\ref{gex}) into (\ref{lxy-lyx}) and obtain
\begin{align}\label{l}
&3G_1'(y) V_1'+3(G_2'(y)-xG_1''(y))V_2'+6(A_{112}y-A_{202}y^2)V_1'V_2'\nonumber\\
&+(x G_1'(y)-G_2(y))(V_1''-V_2'')+6(-A_{013}+A_{103}y-2A_{112}x+4A_{202}xy)(V_2')^2\nonumber\\
&+2x(A_{112}y-A_{202}y^2) V_2'V_1''+8(A_{202} x y^2-A_{112}xy)V_2'V_2''\nonumber\\
&+\frac{\hbar^2}{4}\bigg( (5A_{112}-10A_{202}y)V_1^{(3)}+(5A_{103}+10A_{202}x)V_2^{(3)}\nonumber\\
&+(2A_{202}xy-A_{112}x+A_{103}y-A_{013})(V_2^{(4)}-V_1^{(4)})\bigg)=0.
\end{align}
So far we have identified possible forms of the potential $V_1(x)$ in the case when the linear equations (\ref{**(1&2)}) for $V_2(y)$ are satisfied trivially. Now we shall consider the two classes of potentials $V_1^a,$ and $V_1^b$ separately and obtain nonlinear ODEs for $V_2(y)$. Our main tool for solving these nonlinear ODEs will be singularity analysis. More precisely, we will show that these equations always pass the Painlev\'e test. The same was true in the case of third order integrals of motion. It was shown that the ODEs actually have the Painlev\'e property and they were solved in terms of known Painlev\'e transcendents, or elliptic functions \cite{Gravel:G3, Gravel:GP}. We will now show that the same is true in this case.\\
We define the function
\begin{align}\label{difW}
W(y)=\int{V_2 dy},
\end{align}
and derive ODEs for $W(y)$. Since the potential $V_2(y)$ is defined up to a constant, two integrals $W_1(y)$ and $W_2(y)$ will be considered equivalent if they satisfy
\begin{equation}\label{Weq}
W_2(y)=W_1(y)+\alpha y+\beta; \; \alpha, \beta \in \mathbb{R}
\end{equation}
The ODEs for $W(y)$ will a priori be fourth order nonlinear ones but we will always be able to integrate them once. 
\subsubsection{The potential $V_1^{(b)}(x)=\dfrac{c_{-2}}{x^2}+c_1 x+c_2 x^2, \; c_{-2} \neq 0$}

The potential $V_1^{(b)}$ provides interesting results. It occurs in cases I, and IIb of (\ref{classification}). Solving (\ref{g_3y}) and (\ref{l}) and using (\ref{trivial}) we obtain 
\begin{align}
G_1(y)=&2 y(y A_{202}-A_{112})W'+(2 y A_{202}-A_{112})W-\frac{2}{3} a y^4 A_{202}+\frac{4}{3} a y^3 A_{112}+a_2 y^2+a_1y,\nonumber\\
\end{align}
where $W(y)$ is defined in  (\ref{difW}) and moreover we obtain $ c_1=G_2(y)=G_3(y)=0.$ The function $W(y)$ satisfies the ODE
\begin{align}\label{Vb}
&\frac{1}{4}\hbar ^2(2 A_{202}y- A_{112}) W^{(4)}+2 \hbar ^2 A_{202} W^{(3)}-3(2 A_{202}y- A_{112}) W' W''\nonumber\\
&-2 A_{202} W W''+(\frac{8}{3} c_2 A_{202}y^3 -4 c_2 A_{112}y^2-2 a_2 y-a_1) W''-8 A_{202}W'^2\nonumber\\
&+4(4c_2 A_{202}y^2-4c_2 A_{112}y-a_2) W'+8c_2(2 A_{202}y- A_{112}) W\nonumber\\
&-\frac{16}{3} c_2^2A_{202} y^4+\frac{32}{3} c_2^2  A_{112}y^3+8 a_2 c_2 y^2+8 a_1 c_2 y+k=0,
\end{align}
where $k$ is an integration constant.\\
Case I.  $A_{202} \neq 0, A_{112}=0; Y_L=A_{202}\{L_3^2,p_2^2\}.$\\
Let $A_{202}=1.$
From (\ref{Vb}) and (\ref{Weq}) we obtain
\begin{align}\label{VbII4}
&\frac{1}{2}\hbar ^2 yW^{(4)}+2\hbar^2W^{(3)}-6yW'W''-4WW''+\frac{8}{3}c_2y^3 W''-8W'^2+16c_2y^2W'\nonumber\\
&+16c_2yW-\frac{16}{3}c_2^2y^4+k_1=0,
\end{align}
integrating once we get
\begin{align}\label{VbII3}
&\hbar ^2y ^2 W^{(3)}+2\hbar^2 yW''-6y^2W'^2-4yWW'+(\frac{16}{3}c_2y^4-2\hbar ^2)W'+2W^2+\frac{32}{3}c_2y^3 W\nonumber\\
&-\frac{16}{9}c_2^2y^6+k_1y^2+k_2=0.\nonumber\\
\end{align}
The equation (\ref{VbII3}) passes the Painlev\'e test. Substituting the Laurent series (\ref{laurent}) into (\ref{VbII3}), we find $p=-1$. The resonances are $r=1,$ and $r=6,$ and we obtain  $d_0=-\hbar ^2$. The constants $d_1$ and $d_6$ are arbitrary, as they should be. We now proceed to integrate (\ref{VbII3}).\\
By the following transformation
$$Y=y^2,\; U(Y)=-\frac{y}{2\hbar ^2}W(y)+\frac{c_2}{6 \hbar ^2} y^4+\frac{1}{16},$$
we transform (\ref{VbII3}) to
\begin{align}\label{chazyI}
Y^2U^{(3)}=-2(U'(3YU'-2U)-\frac{c_2}{\hbar ^2} Y (Y U'-U)+k_3Y+k_4)-YU'',
\end{align}
where
$k_3=\frac{-2 k_1-12 c_2 \hbar ^2}{64 \hbar ^4},\;k_4=\frac{- k_2}{32 \hbar ^4}.$
The equation (\ref{chazyI}) is a special case of the Chazy class I equation. 
It admits the first integral
\begin{align}\label{SDIb}
Y^2U''^2=-4(U'^2(YU'-U)-\frac{c_2}{2\hbar ^2}(YU'-U)^2+k_3(YU'-U)+k_4U'+k_5),
\end{align}
where $k_5$ is the integration constant. The equation is the canonical form SD-I.b in \cite[page 65-73]{Cosgrove:SD}.
When $c_2$ and $k_3$ are both nonzero the solution is
\begin{align}
U=&\frac{1}{4}(\frac{1}{P_{5}}(\frac{YP_5'}{P_{5}-1}-P_{5})^2-(1-\sqrt{2\alpha})^2(P_{5}-1)-2\beta \frac{P_{5}-1}{P_{5}}+\gamma Y \frac{P_{5}+1}{P_{5}-1}+2 \delta \frac{Y^2P_{5}}{(P_{5}-1)^2}),\nonumber\\
U'=&-\frac{Y}{4P_{5}(P_{5}-1)}(P_{5}'-\sqrt{2\alpha} \frac{P_{5}(P_{5}-1)}{Y})^2-\frac{\beta}{2Y}\frac{P_{5}-1}{P_{5}}-\frac{1}{2}\delta Y \frac{P_{5}}{P_{5}-1}-\frac{1}{4}\gamma,\nonumber\\
\end{align}
where $P_5=P_5(Y); Y=y^2,$ satisfies the fifth Painlev\'e equation
$$P_5''=(\frac{1}{2P_5}+\frac{1}{P_5-1})P_5'^2-\frac{1}{Y}P_5'+\frac{(P_5-1)^2}{Y^2}(\alpha P_5+\frac{\beta}{P_5})+\gamma \frac{P_5}{Y}+\delta \frac{P_5(P_5+1)}{P_5-1},$$
with
$$c_2=-\hbar ^2\delta,\; k_3=-\frac{1}{4}(\frac{1}{4}\gamma ^2+2\beta \delta -\delta (1-\sqrt{2 \alpha})^2),\; k_4=-\frac{1}{4}(\beta \gamma+\frac{1}{2}\gamma (1-\sqrt{2\alpha})^2),$$
$$k_5=-\frac{1}{32}(\gamma ^2((1-\sqrt{2 \alpha})^2-2\beta)-\delta((1-\sqrt{2 \alpha})^2+2\beta)^2).$$
The solution for the potential up to a constant is
\begin{align}\label{VbP5}
V(x,y)=&\dfrac{c_{-2}}{x^2}-\delta \hbar ^2 (x^2+y^2)+\hbar ^2 \big( \frac{\gamma}{P_5-1}+\frac{1}{y^2}(P_5-1)(\sqrt{2\alpha }+\alpha(2P_5-1)+\frac{\beta}{P_5})\nonumber\\
&+y^2(\frac{P_5'^2}{2 P_5}+\delta P_5 )\frac{(2P_5-1)}{(P_5-1)^2}-\frac{P_5'}{P_5-1}-2\sqrt{2\alpha}P_5'\big)+\frac{3\hbar^2}{8y^2}.\nonumber\\
\end{align}
And we have
\begin{align}
g_1(x,y)=&2 y^2W'+2 y W+\frac{2}{3}\hbar ^2\delta y^4,\;g_2(x,y)=-x(6yW'+2W+\frac{8}{3}\hbar ^2\delta y^3),\nonumber\\
g_{3}(x,y)&=4x^2W'+2\hbar ^2\delta x^2y^2+2c_{-2}\frac{y^2}{x^2},\nonumber\\
l(x,y)&= \hbar^2 x^2(\frac{1}{4}y W^{(4)}+W^{(3)})-x^2( 3y W'+W )W''-(\frac{4}{3}\hbar ^2\delta x^2y^3+\frac{3 \hbar^2}{2}y)W''\nonumber\\
&+(4(\dfrac{c_{-2}}{x^2}-\hbar ^2\delta x^2)y^2-3 \hbar ^2)W'+4y(\dfrac{c_{-2}}{x^2}-\hbar ^2\delta x^2)W+\dfrac{4c_{-2}}{3x^2}\hbar ^2\delta  y^4 \nonumber\\
&-2 \hbar ^2\delta x^2(\frac{2}{3} \hbar ^2\delta y^4- \hbar^2)-2 \hbar ^4\delta y^2.\nonumber\\
\end{align}
The solution of (\ref{SDIb}) when $c_2=0$ is 
\begin{align}
U=&\frac{1}{4}(\frac{1}{P^2}(YP'-P)^2-\frac{1}{16} \alpha P^2-\frac{1}{8}(\beta+2\sqrt{\alpha})P+\frac{1}{8P}\gamma Y+\frac{1}{16 P^2}\delta Y^2),\nonumber\\
U'=&-\frac{1}{4}\sqrt{\alpha}P'-\frac{1}{8Y}(\alpha P^2+\beta P),\nonumber\\
k_3&=\frac{1}{64}\alpha \delta,\; k_4=-\frac{1}{64}\gamma (\beta +2\sqrt{\alpha}),\;k_5=-\frac{1}{1024}(\alpha \gamma ^2-\delta (\beta +2\sqrt{\alpha})^2),\nonumber\\
\end{align}
where $P(Y)=yP_{3}(y)$, and $P_3$ satisfies the third Painlev\'e equation
$$P_3''=\frac{P_3'^2}{P_3}-\frac{P_3'}{y}+\alpha P_3^3+\frac{\beta P_3^2+\gamma}{y}+\frac{\delta}{P_3}.$$
The solution for the potential is
\begin{align}\label{VbP3}
V(x,y)=\dfrac{c_{-2}}{x^2}+ \frac{\hbar ^2}{2}(\sqrt{\alpha } P_3'+\frac{3}{4} \alpha  P_3^2+\frac{\delta }{4P_3^2}+\frac{\beta P_3}{2y}+\frac{\gamma}{2 P_3y}-\frac{ P_3'}{2y P_3}+\frac{P_3'^2}{4 P_3^2}).
\end{align}
And we have
\begin{align}
g_1(x,y)=&2 y^2W'+2 y W,\;g_2(x,y)=-6xyW'-2xW,\nonumber\\
g_{3}(x,y)&=4x^2W'+2c_{-2}\frac{y^2}{x^2},\nonumber\\
l(x,y)&= \hbar^2 x^2(\frac{1}{4}y W^{(4)}+W^{(3)})-x^2( 3y W'+W )W''-\frac{3}{2} \hbar^2 yW''+(4\dfrac{c_{-2}}{x^2}y^2-3 \hbar^2)W'+4\dfrac{c_{-2}}{x^2}yW.\nonumber\\
\end{align}
Case IIb. $A_{202}=0, A_{112} \neq 0; Y_L=A_{112}\{L_3,p_1p_2^2\}.$\\
Let $A_{112}=1.$ In this case,
\begin{align}
g_1(x,y)=&-2 yW'-W+\frac{4}{3} c_2 y^3+a_2 y^2,\;g_2(x,y)=3xW'-4c_2xy^2-2a_2xy,\nonumber\\
g_{3}(x,y)&=2c_2x^2y+a_2x^2-2c_{-2}\frac{y}{x^2},\nonumber\\
l(x,y)&= -\frac{1}{8} \hbar^2 x^2 W^{(4)}+\frac{3}{2} x^2 W'W''-(2 c_2 x^2y^2+a_2 x^2 y-\frac{3 }{4}\hbar ^2)W''-2(2c_{-2} \frac{y }{x^2}+2c_2 x^2 y)W'\nonumber\\
&-2(\dfrac{c_{-2}}{x^2}+c_2 x^2 )W+2x^2(\frac{4}{3}c_2^2 y^3+a_2c_2 y^2)+2\dfrac{c_{-2}}{x^2}(\frac{4}{3}c_2y^3+ a_2 y^2)-2 c_2 \hbar^2y.\nonumber\\
\end{align}
Integrating the equation (\ref{Vb}) we get 
\begin{equation}\label{VbI}
\hbar ^2W^{(3)}-6W'^2+8(2c_2y^2+a_2y)W'+8(4c_2y+a_2)W-\frac{32}{3}c_2^2y^4-\frac{32}{3}c_2a_2y^3+k_1y+k_2=0.
\end{equation}
The equation (\ref{VbI}) passes the Painlev\'e test. Substituting the Laurent series (\ref{laurent}) into (\ref{VbI}), we obtain $p=-1$. The resonances are $r=1,$ and $r=6,$ and  $d_0=-\hbar ^2$. The constants $d_1$ and $d_6$ are arbitrary. By an appropriate linear transformation of the form
$$Y=\lambda_1 y+\lambda_2, \; U(Y)=\lambda_3 W(y)+\mu(y),$$
we transform the equation (\ref{VbI}) into a special case of the canonical form for the Chazy class I.  The general form of the equation is
\begin{align}\label{VbU}
U^{(3)}=-2 (3 U'^2+2 k_3 Y (Y U'-U)+k_4 (2 Y U'-U)+2 k_5 U'+k_6Y+k_7),
\end{align}
Depending on the choice of $\lambda_1,\lambda_2, \lambda_3$ and $\mu$, the parameters $k_3,k_4,k_5,k_6,$ and $k_7$ get different values, and the first integral of the equation (\ref{VbU}) with respect to $Y$ corresponds to one of the four canonical subcases, listed below.
For $c_2 \neq 0, k_3=-1, k_4=k_5=k_6=0,$ we get equation $SD-I.c$:\\
\begin{align}\textbf{•}
U''^2=-4(U'^3-(YU'-U)^2+k_7 U'+k_8),\nonumber\\
\end{align}
where $k_8$ is the integration constant. The solution for the equation $SD-I.c$ is 
\begin{align}
U=&\frac{1}{8P_{4}}P_{4}'^2-\frac{1}{8}P_{4}^3-\frac{1}{2}Y P_{4}^2-\frac{1}{2}(Y^2-\alpha +\epsilon )P_{4}+\frac{1}{3}(\alpha -\epsilon )Y+\frac{\beta}{4P_{4}},\nonumber\\
U'=&-\frac{1}{2}\epsilon P_4'-\frac{1}{2}P_4^2-YP_4+\frac{1}{3}(\alpha -\epsilon),
\end{align}
where 
$$\epsilon =\pm 1, k_7=-\frac{1}{3}(\alpha -\epsilon)^2-2\beta , k_8=\frac{1}{3}(\alpha -\epsilon)(\beta +\frac{2}{9}(\alpha-\epsilon)^2),$$
and $P_4=P_4(-\sqrt[4]{\frac{8c_2}{\hbar ^2}} y-\frac{a_2}{2\sqrt[4]{2 c_2^3 \hbar ^2}}),$ satisfies the fourth Painlev\'e equation (for arbitrary $\alpha$ and $\beta$)
$$P_{4}''=\frac{P_{4}'^2}{2P_{4}}+\frac{3}{2}P_{4}^3+4YP_{4}^2+2(Y^2-\alpha )P_{4}+\frac{\beta}{P_{4}}.$$
Therefore, the solution for potential is
\begin{align}\label{VbP4}
V(x,y)=&2 a_2y+c_2( x^2+4 y^2)+\dfrac{c_{-2}}{x^2}-\frac{\sqrt[4]{2} a_2 \sqrt{\hbar} P_4}{\sqrt[4]{c_2}}-4 \sqrt[4]{2 c_2^3}
   \sqrt{\hbar}y P_4 +\sqrt{2c_2} \hbar( \epsilon  P_4'+ P_4^2).\nonumber\\
\end{align}
For $a_2 \neq 0, c_2=k_3=k_5=k_6=k_7=0,k_4=\frac{1}{2}$ we obtain equation $SD-I.d$:\\
\begin{align}
U''^2=-4U'^3-2U'(YU'-U)+k_8.\nonumber\\
\end{align}
The solution for the equation $SD-I.d$ is
\begin{align}
U=&\frac{1}{2}(P_2')^2-\frac{1}{2}(P_2^2+\frac{1}{2}Y)^2-(\alpha+\frac{1}{2}\epsilon)P_2,\nonumber\\
U'=&-\frac{1}{2}(\epsilon P_2'+P_2^2+\frac{1}{2}Y),
\end{align}
where $k_8=\frac{1}{4}(\alpha +\frac{1}{2}\epsilon)^2,$ and $P_2(Y)=P_2(-2\sqrt[3]{\frac{a_2}{\hbar^2}}y-\frac{3 k_1}{16 \sqrt[3]{a_2^5 \hbar ^2}}),$ satisfies the second Painlev\'e equation
$$P_2''=2P_2^3+YP_2+\alpha.$$
Therefore, the solution for potential is
\begin{align}\label{VbP2}
V(x,y)=&\dfrac{c_{-2}}{x^2}+2 \sqrt[3]{a_2^2 \hbar ^2}( \epsilon  P_2'+ P_2^2).
\end{align}
For $c_2=a_2=k_3=k_4=k_5=k_7=0, k_6=\frac{1}{2},$ we get equation $SD-I.e$:\\
\begin{align}
U''^2=-4U'^3-2(YU'-U).\nonumber\\
\end{align}
The solution for the equation $SD-I.e$ is
$$U=\frac{1}{2}(P_1')^2-2P_1^3-Y P_1, \quad U'=-P_1.$$
The function $P_1(Y)=P_1(-\sqrt[5]{\frac{k_1}{\hbar ^4}}y-\frac{k_2}{\sqrt[5]{ k_1^4 \hbar ^4}}),$ satisfies the first Painlev\'e equation
$$P_1''=6P_1^2+Y,$$
and we have
\begin{align}\label{VbP1}
V(x,y)=&\dfrac{c_{-2}}{x^2}+\sqrt[5]{k_1^2 \hbar ^2} P_1.
\end{align}
For $c_2=a_2=k_3=k_4=k_5=k_6=0,$ we obtain equation $SD-I.f$:\\
\begin{align}
U''^2=-4(U'^3+k_7U'+k_8).\nonumber\\
\end{align}
The solution for the equation $SD-I.f$ is
$$U=-\int{u}dy +\alpha_1,\quad u=\wp(y-\alpha_2, -4k_7 ,4k_8)$$
where $\alpha_1, \alpha_2$ are integration constants, and $\wp$ is the Weierstrass elliptic function. Thus
\begin{align}\label{Vbwei}
V(x,y)&=\dfrac{c_{-2}}{x^2}+\hbar ^2 \wp .
\end{align}
\\
\subsubsection{The potential $V_1^{(a)}(x)=c_1 x+c_2 x^2; \;c_1c_2=0$}\label{potentialho}

We again define $W(y)$ as in (\ref{difW}). From (\ref{g_3y}) and (\ref{trivial}) we obtain
\begin{align}\label{Gh}
G_1(y)&=2(A_{202}y^2-A_{112}y)W'+(2A_{202}y-A_{112})W-\frac{2}{3}c_2A_{202}y^4+\frac{4}{3}c_2A_{112}y^3+a_2y^2+a_1y,\nonumber\\
G_2(y)&=-3(A_{103}y-A_{013})W'-A_{103}W-c_1A_{202}y^3+\frac{1}{3}c_2A_{103}y^3-\frac{3}{2}c_1A_{112}y^2-c_2A_{013}y^2+b_1y+b_0,\nonumber\\
G_3(y)&=-c_1y(\frac{1}{2}A_{103}y-A_{013}).\nonumber\\
\end{align}
Substituting $G_1,G_2,G_3$ in (\ref{l}) and integrating it with respect to $y$, we get
$$K_1x+K_2=0,$$
where
\begin{align}\label{K1K2}
K_1&=\hbar ^2(A_{112}-2A_{202}y)W^{(4)}-8\hbar^2 A_{202}W^{(3)}+12(2A_{202}y-A_{112})W'W''+8A_{202}WW''\nonumber\\
&-4(\frac{8}{3}c_2A_{202}y^3-4c_2A_{112}y^2-2a_2y-a_1)W''+32A_{202}W'^2-16(4c_2A_{202}y^2-4c_2A_{112}y-a_2)W'\nonumber\\
&-32c_2(2A_{202}y-A_{112})W+\frac{64}{3}c_2^2A_{202}y^4-\frac{128}{3}c_2^2A_{112}y^3 -32a_2c_2y^2 -32a_1c_2y+k_1,\nonumber\\
K_2&=\hbar ^2(A_{013}-A_{103}y)W^{(4)}-4\hbar^2A_{103}W^{(3)}+12(A_{103}y-A_{013})W'W''+4A_{103}WW''\nonumber\\
&-(4c_1A_{202}y^3+\frac{4}{3}c_2A_{103}y^3-6c_1A_{112}y^2-4c_2A_{013}y^2+4b_1y+4b_0)W''+16A_{103}W'^2\nonumber\\
&-8(3c_1A_{202}y^2+c_2A_{103}y^2-2c_2A_{013}y-3c_1A_{112}y+b_1)W'\nonumber\\
&-4(6c_1A_{202}y+2c_2A_{103}y-3c_1A_{112}-2c_2A_{013})W+\frac{2}{3}c_2^2A_{103}y^4-\frac{8}{3}c_2^2A_{013}y^3+4c_2b_1y^2\nonumber\\
&-12a_2c_1y^2+8c_2b_0y-12a_1c_1y+k_2,\nonumber\\
\end{align}
and we must have $K_1=0,\; K_2=0.$ In general the two ODEs  in (\ref{K1K2}) are not compatible and we will analyze their compatibility conditions. A crucial role is played by the matrix
$$A = 
 \begin{pmatrix}
  A_{202} & A_{112}  \\
  A_{103} & A_{013}  \\
 \end{pmatrix}.
 $$
For the integral (\ref{V2trivially}) to exist the rank of $A$ must be $1,$ or $2$. Let us analyze different possibilities.\\
1. rank$(A)=1, \;A_{112}=A_{202}=0.$ In this case $K_1=0$ reduces to a linear second order ODE for $W(y)$;
\begin{align}\label{1W}
(a_1+2a_2y)W''+ 4a_2 W'-8 y a_1 c_2 - 8a_2c_2 y^2+\frac{k_1}{4}=0.
\end{align}
For $(a_2,a_1) \neq (0,0),$ equation (\ref{1W}) together with $K_2=0$ leads to the elementary potentials that allow second order integrals of motion. They were already discussed in \cite{Fris:Fri}. Of more interest is the case when we also have $a_1=a_2=0,$ so (\ref{1W}) is satisfied identically, and $K_2=0$ reduces to
\begin{align}\label{K2}
&\hbar ^2(A_{013}-y A_{103})W^{(4)}-4 \hbar ^2 A_{103} W^{(3)}+4(A_{103} W-\frac{1}{3} c_2 y^3 A_{103}+ c_2 y^2 A_{013}- b_1 y-b_0)W''+\nonumber\\
&12( y A_{103}- A_{013})W' W''+16A_{103} W'^2-8( c_2 y^2A_{103}-2 c_2 y A_{013}+ b_1)W'+8c_2(A_{013}- y A_{103})W+\nonumber\\
&4 b_1 c_2 y^2+8 b_0 c_2 y+\frac{2}{3} c_2^2 y^4A_{103}-\frac{8}{3} c_2^2 y^3 A_{013}+k_2=0.\nonumber\\
\end{align}
Thus, we have one 4th order nonlinear ODE to solve and we must distinguish two cases, according to (\ref{classification}).\\
Case IIa. $A_{103}\neq 0, A_{202}=A_{112}=A_{013}=0 ; Y_L=A_{103}\{L_3,p_2^3\}.$\\
Setting $A_{103}=1,$ we obtain\\
\begin{align}
g_1(x,y)&=0,\;g_2(x,y)=-3yW'-W+\frac{1}{3}c_2y^3,\;g_{3}(x,y)=4xW'-c_2xy^2-\frac{1}{2}c_1y^2,\nonumber\\
l(x,y)=&\frac{1}{4} \hbar^2 x (y W^{(4)}+4 W^{(3)})-3 x y W'^2-xWW'+\frac{1}{3} c_2 x y^3 W''-c_1 y^2 W'-c_1 y W-\frac{1}{2} \hbar^2 c_2 x.\nonumber\\
\end{align}
From (\ref{K2}) we have
\begin{align}\label{W103}
\hbar^2y W^{(4)}+4\hbar^2W^{(3)}-12yW'W''-4WW''+\frac{4}{3}c_2y^3W''-16W'^2+8c_2y^2W'+8c_2yW-\frac{2}{3}c_2^2y^4+k=0.\nonumber\\
\end{align}
This equation is the same type of equation as (\ref{VbII4}), (with slightly different parameters, and $c_2$ in (\ref{VbII4}) is replaced by $\frac{c_2}{4}$), and has solutions expressed in terms of the fifth and third Painlev\'e transcendents. For $c_2 \neq 0,$ we have
\begin{align}\label{VaP5}
V(x,y)=&-\delta \hbar ^2  (4x^2+y^2)+\hbar ^2 \big( \frac{\gamma}{P_5(y^2)-1}+\frac{1}{y^2}(P_5(y^2)-1)(\sqrt{2\alpha }+\alpha(2P_5(y^2)-1)+\frac{\beta}{P_5(y^2)})\nonumber\\
&+y^2(\frac{P_5'^2(y^2)}{2 P_5(y^2)}+\delta P_5(y^2))\frac{(2P_5(y^2)-1)}{(P_5(y^2)-1)^2}-\frac{P_5'(y^2)}{P_5(y^2)-1}-2\sqrt{2\alpha}P_5'(y^2)\big)+\frac{3\hbar^2}{8y^2}, \nonumber\\
\end{align}
and for $c_2=0$,
\begin{align}\label{VaP3}
V(x,y)=c_1x+\frac{\hbar ^2}{2}(\sqrt{\alpha } P_3'(y)+\frac{3}{4} \alpha  (P_3(y))^2+\frac{\delta }{4P_3^2(y)}+\frac{\beta P_3(y)}{2y}+\frac{\gamma}{2y P_3(y)}-\frac{ P_3'(y)}{2y P_3(y)}+\frac{P_3'^2(y)}{4 P_3^2(y)}).
\end{align}
Case III. $A_{202}=A_{112}=A_{103}=0, A_{013} \neq 0; Y_L=2A_{013}p_1p_2^3.$\\
We set $A_{013}=1.$
\begin{align}
g_1(x,y)&=0,\;g_2(x,y)=3W'-c_2y^2+b_1y,\;g_{3}(x,y)=2c_2xy+c_1y-b_1x,\nonumber\\
l(x,y)=&-\frac{1}{4} \hbar^2 x W^{(4)}+3 x W'W''+(b_1 x y-c_2 xy^2)W''+2c_1 y W'+c_1 W+\frac{1}{2} b_1 c_1 y^2.\nonumber\\
\end{align}
Integrating (\ref{K2}), we get
\begin{align}\label{W013}
\hbar^2W^{(3)}-6W'^2+4(c_2y^2-b_1y)W'+(8c_2y-4b_1)W-\frac{2}{3}c_2^2y^4+\frac{4}{3}b_1c_2y^3+k_2y+k_3=0,
\end{align}
which is the same type of equation as (\ref{VbI}), (with slightly different parameters, and $c_2$ in (\ref{VbI}) is replaced by $\frac{c_2}{4}$ ) and can be solved in terms of the fourth, second and first Painlev\'e transcendents and elliptic functions. Depending on the values of the parameters in (\ref{W013}) and following the procedure after (\ref{VbU}), we obtain the following potentials.\\
When $c_2 \neq 0,$ $c_1 =0,$ and the potential is
\begin{align}\label{VaP4}
V(x,y)=&-b_1y+c_2( x^2+y^2)-\frac{b_1 \sqrt{\hbar} P_4}{\sqrt[4]{2c_2}}- \sqrt[4]{8c_2^3 \hbar ^2}y P_4 +\sqrt{\frac{c_2}{2}} \hbar( \epsilon  P_4'+ P_4^2),\nonumber\\
\end{align}
where $\epsilon =\pm 1,$ and $P_4=P_4(-\sqrt[4]{\frac{2c_2}{\hbar ^2}} y+\frac{b_1}{\sqrt[4]{2^3 c_2^3 \hbar ^2}}),$ satisfies the fourth Painlev\'e equation.\\
When $c_2=0, b_1 \neq 0,$ the solutions are
\begin{align}\label{VaP2}
V(x,y)=&c_1x+ \sqrt[3]{2b_1^2 \hbar ^2}( \epsilon  P_2'+ P_2^2),
\end{align}
where $P_2=P_2(\sqrt[3]{\frac{4b_1}{\hbar^2}}y+\frac{3 k_2}{2 \sqrt[3]{4b_1^5 \hbar ^2}}),$ satisfies the second Painlev\'e equation.\\
For $c_2=b_1=0, k_2 \neq 0,$ the potential is
\begin{align}\label{VaP1}
V(x,y)=&c_1x+\sqrt[5]{ k _2^2 \hbar^2} P_1,
\end{align}
for $P_1=P_1(-\sqrt[5]{\frac{k_2}{\hbar^4}}y-\frac{k_3}{\sqrt[5]{k _2^4 \hbar^4}}),$ satisfying the first Painlev\'e equation.
and finally, for $c_2=b_1=k_2=0,$ we are left with
\begin{align}\label{Vawei}
V(x,y)=&c_1x +\hbar ^2 \wp ,
\end{align}
where $\wp$ is the Weierstrass elliptic function.\\\\
2. rank$(A)=1, \;A_{013}=A_{103}=0.$ In this case $K_2=0$ reduces to a linear second order ODE
\begin{align}\label{K22}
&(4c_1A_{202}y^3-6c_1A_{112}y^2+4b_1y+4b_0)W''+8(3c_1A_{202}y^2-3c_1A_{112}y+b_1)W'+4(6c_1A_{202}y-3c_1A_{112})W\nonumber\\
&-4c_2b_1y^2+12a_2c_1y^2-8c_2b_0y+12a_1c_1y-k_2=0.\nonumber\\
\end{align}
Since at least one of $A_{112}$ and $A_{202}$ must be nonvanishing, (\ref{K22}) leads to elementary potentials (unless it satisfied trivially). Equation (\ref{K22}) is satisfied trivially if $c_1=b_1=b_0=0.$ We are left with one fourth order nonlinear ODE, $K_1=0.$ In view of (\ref{classification}) two cases must be considered.\\\\
Case I. $A_{202}\neq 0, A_{112}=A_{103}=A_{013}=0; Y_L=A_{202}\{L_3^2,p_2^2\}$.\\
In this case, we have
\begin{align}
g_1(x,y)=&2 y^2W'+2yW-\frac{2}{3} c_2 y^4,\;g_2(x,y)=-6xyW'-2xW+\frac{8}{3}c_2xy^3,\; g_{3}(x,y)=x^2W'-2c_2x^2y^2,\nonumber\\
l(x,y)&=\hbar^2 x^2(\frac{1}{4}  y W^{(4)}+W^{(3)})-x^2(3y W'+W)W''+(\frac{4}{3} c_2 x^2 y^3-\frac{3 \hbar^2}{2}y)W''+(4c_2x^2y^2-3 \hbar^2)W'\nonumber\\
&+4c_2 x^2y W-c_2^2(\frac{4}{3}x^2+\frac{1}{4})y^4+2 c_2 \hbar^2( y^2-x^2).\nonumber\\
\end{align}
and
\begin{align}\label{K1202}
&\hbar ^2 yW^{(4)}+4\hbar^2W^{(3)}-12yW'W''-4WW''+\frac{16}{3}c_2y^3W''-16W'^2+32c_2y^2W'+32c_2yW-\frac{32}{3}c_2^2y^4+k=0,\nonumber\\
\end{align}
which is exactly the same equation as (\ref{VbII4}) and hence has the same solutions expressed in terms of the fifth and third Painlev\'e transcendents.\\
Case IIb. $A_{112} \neq 0, A_{202}=A_{103}=A_{013}=0; Y_L=A_{112}\{L_3,p_1p_2^2\}.$

\begin{align}
g_1(x,y)&=-2 yW'-W+\frac{4}{3} c_2 y^3+a_2 y^2,\;g_2(x,y)=3xW'-4c_2xy^2-2a_2xy,\; g_{3}(x,y)=2c_2x^2y+a_2x^2,\nonumber\\
l(x,y)=&-\frac{1}{8} \hbar^2 x^2 W^{(4)}+\frac{3}{2} x^2 W'W''+(-2 c_2 x^2 y^2+\frac{3 \hbar^2}{4})W''-4c_2x^2yW'-2c_2 x^2 W+\frac{8}{3} c_2^2 x^2 y^3-2 c_2 \hbar^2 y.\nonumber\\
\end{align}
Integrating $K_1=0$ once we obtain
\begin{equation}\label{IIb2}
\hbar ^2W^{(3)}-6W'^2+8(2c_2y^2+a_2y)W'+8(4c_2y+a_2)W-\frac{32}{3}c_2^2y^4-\frac{32}{3}c_2a_2y^3+k_3y+k_4=0,
\end{equation}
which is the same equation as (\ref{VbI})  and is solved in terms of the fourth, second and first Painlev\'e transcendents and elliptic function.\\
3. rank$(A)=2$. Both $K_1=0,$ and $K_2=0,$ are satisfied nontrivially.\\
Case I. $A_{202}\neq 0, A_{013} \neq 0; A_{112}=A_{103}=0; Y=A_{202}\{L_3^2,p_2^2\}+2A_{013} p_1p_2^3$.\\
Let us set $A_{202}=1, A_{013}=\alpha,$ with $\alpha \neq 0.$ In this case, both equations in (\ref{K1K2}) can be integrated once and we obtain two third order equations
\begin{align}\label{intK1}
&\hbar ^2 y^2 W^{(3)}+2\hbar ^2 y W''-6 y^2 W'^2+(\frac{16 c_2}{3}y^4-4 a_2 y^2-2 a_1 y-2 \hbar ^2-4y W)W'+2 W^2\nonumber\\
&+(2a_1+\frac{32 c_2 }{3}y^3)W-(\frac{16}{9} c_2^2 y^4-4 a_2 c_2 y^2-\frac{16}{3} a_1 c_2y+\frac{k_1}{4})y^2+k_3=0,\nonumber\\
\end{align}
\begin{align}\label{intK2}
&\alpha \hbar ^2W^{(3)}-6\alpha W'^2-4(c_1y^3-\alpha c_2y^2+b_1y+b_0)W'-4(3c_1y^2-2c_2\alpha y+b_1)W\nonumber\\
&-\frac{2}{3}c_2^2\alpha y^4+4(\frac{1}{3}c_2b_1-c_1a_2)y^3-2(3c_1a_1-2c_2b_0)y^2+ k_2y+k_4=0,\nonumber\\
\end{align}
where $k_3$ and $k_4$ are integration constants. Eliminating third order derivatives between (\ref{intK1}) and (\ref{intK2}), we obtain a second order ODE. This equation admits a first integral,
\begin{align}\label{riccati1}
&\alpha  \hbar ^2 W'-\alpha  W^2- (\alpha  a_1-2( b_0- \alpha  a_2)y-2b_1 y^2-\frac{2}{3}\alpha c_2 y^3-2 c_1 y^4)W-\frac{1}{9} \alpha  c_2^2 y^6+\frac{1}{6}(3 a_2 c_1-b_1 c_2)y^5\nonumber\\
&+(\frac{2}{3} (\alpha  a_2 c_2-b_0c_2)+a_1 c_1)y^4+(\frac{4}{3} \alpha  a_1c_2-\frac{k_2}{4})y^3-\frac{1}{8}(\alpha  k_1+4 k_4) y^2+k_5y-\alpha \frac{k_3}{2}=0,\nonumber\\
\end{align}
where $k_5$ is an integration constant. Equation (\ref{riccati1}) is a Riccati equation and can be linearized by a Cole-Hopf transformation.
Setting $W=-\hbar ^2 \frac{U'}{U},$ we get the following linear ODE
\begin{align}
&\alpha \hbar ^4 U''(y)+\hbar^2(2 c_1 y^4+\frac{2}{3} \alpha  c_2 y^3+2 b_1 y^2-2 \alpha  a_2 y+2 b_0y-\alpha  a_1) U'(y)+(\frac{1}{9} \alpha  c_2^2 y^6- (\frac{a_2 c_1}{2}-\frac{b_1c_2}{6})y^5\nonumber\\
&-(\frac{2}{3} \alpha  a_2 c_2+a_1c_1-\frac{2 b_0 c_2}{3})y^4-(\frac{4}{3}\alpha  a_1 c_2-\frac{k_2}{4})y^3+ \frac{1}{8}(4k_3+\alpha  k_1)y^2+\frac{k_5}{2\hbar ^2}y+\frac{\alpha k_3}{2})U(y)=0.\nonumber\\
\end{align}
Consequently, in this case we do not obtain any exotic potential.\\
Case II. $A_{202}=0, A_{112} \neq 0, A_{103} \neq 0; Y_L=A_{112}\{L_3,p_1p_2^2\}+A_{103} \{L_3,p_2^3\}$.\\
Same as the previous case, we can again integrate the equations in (\ref{K1K2}), and if we apply the same procedure we generate another Riccati equation
\begin{align}
&\alpha  \hbar ^2 W'-\alpha  W^2+ (2b_0+2(b_1-\alpha  a_1)y-(3c_1+4\alpha a_2)y^2-\frac{22}{3}\alpha c_2 y^3)W+\frac{19}{18} \alpha c_2^2 y^6+(\frac{4}{3}\alpha a_2 c_2+\frac{1}{2}c_1 c_2)y^5\nonumber\\
&+(\frac{8}{3}\alpha a_1c_2-\frac{1}{6} b_1 c_2+\frac{1}{2}a_2c_1)y^4+(a_1c_1-\frac{2}{3} b_0 c_2-\frac{1}{4}\alpha k_1)y^3-(\frac{1}{4} k_2+k_3) y^2+k_4=0,\nonumber\\
\end{align}
where $k_3,$ and $k_4$ are constants of integration. Again it can be linearized by a Cole-Hopf transformation.
\subsection{LINEAR EQUATIONS FOR $V_1$ SATISFIED TRIVIALLY}

In this case, (\ref{**}) are valid, and (\ref{*}) not. The leading-order term for the nontrivial fourth order integral has the form
\begin{align}\label{IV2}
Y_L=A_{220}\{L_{3}^2, p_{1}^2\}+A_{130}\{L_{3},p_{1}^3\}+A_{121}\{L_{3},p_{1}^2 p_{2}\}+2A_{031}p_{1}^3 p_{2}.
\end{align}
Let us classify the integrals (\ref{IV2}) under translations. The three classes are
\begin{align}
&(i)A_{220} \neq 0, A_{121}=A_{130}=0.\nonumber\\
&(ii)A_{220}=0, A_{121}^2+A_{130}^2 \neq 0, A_{031}=0.\nonumber\\
&(iii)A_{220}=A_{121}=A_{130}=0, A_{031}\neq 0.
\end{align}
Since we can just adapt the results from the section $4.2$ to this case, we will not consider it separately. The results are obtained by interchanging $x \leftrightarrow y, \; (A_{202},A_{112},A_{103},A_{013}) \leftrightarrow (A_{220},A_{121},A_{130},A_{031}).$
\section{Classical analogs of the quantum exotic potentials}

In the classical case, we are dealing with the classical limit ($\hbar \to 0$) of the determining equations (\ref{det1}) and (\ref{l1}) and therefore the compatibility condition (\ref{V}) and (\ref{lxy-lyx}). The equations (\ref{det1}) and (\ref{V}) are actually the same in the classical and quantum case. We continue our investigation for the classical potentials followed by the classifications of the integrals in (\ref{classification}). Here we present the results briefly for each cases.\\
Integrating the classical analog of the equations (\ref{VbII4}), (\ref{W103}) and (\ref{K1202}), we get
\begin{align}\label{clas1}
3y^2W'^2+(2yW-\frac{2}{3}\lambda y^4)W'-W^2-\frac{4}{3}\lambda y^3W+\frac{1}{18}\lambda ^2y^6+k_1y^2+k_2=0,\nonumber\\
\end{align}
where $\lambda=c_2,4c_2$ respectively for $Y_L=L_3 p_{2}^3,$ and  $Y_L=L_3^2 p_2^2$.\\
The classical analog of the equations (\ref{VbI}), (\ref{W013}), and (\ref{IIb2}) is
\begin{align}\label{clas2}
W'^2+\frac{2}{3}y(k_1-\lambda y)W'+\frac{2}{3}(k_1-2\lambda y)W+\frac{1}{9}\lambda ^2y^4-\frac{2}{9}\lambda k_1 y^3+k_2y+k_3=0,\nonumber\\
\end{align}
where $\lambda=c_2,4c_2$ respectively for $Y_L=p_1 p_{2}^3,$ and  $Y_L=L_3 p_1 p_2^2$.\\
Equations (\ref{clas1}) and (\ref{clas2}) are special cases of equation (\ref{1Nd}). They do not satisfy the conditions in the Fuchs' theorem, (Theorem1.1, \cite[page 80]{Chalkley:fuchs}, proof in \cite[page 304-311]{Ince:ode}), hence do not have the Painlev\'e property. They will be further investigated in Part II of this project.
\section{SUMMARY OF RESULTS AND FUTURE OUTLOOK}

\subsection{ Quantum potentials}
The list of exotic superintegrable quantum potentials in quantum case that admit one second order and one fourth order integral is given below. We also give their fourth order integrals by listing the leading terms $Y_L$ and the functions $g_i(x,y); i=1,2,3;$ and $l(x,y).$ Each of the exotic potentials has a non-exotic part that comes from $V_1(x)$. By construction $V_2(y)$ is exotic, however in 4 cases a non-exotic part proportional to $y^2$ splits off from $V_2(y)$ and can be combined with an $x^2$ term in $V_1(x)$. We order the final list below in such a manner that the first two potentials are isotropic harmonic oscillators (possibly with an additional $\dfrac{1}{x^2}$ term) with an added exotic part. The next two are $2:1$ anisotropic harmonic oscillators, plus an exotic part (in $y$).\\
Based on previous experience (see Marquette \cite{Marquette:Rational, Marquette:Pain, Marquette:3order}) we expect these harmonic terms to determine the bound state spectrum. The remaining $8$ cases have either $\dfrac{a}{x^2}$ or $c_1x$ as their non-exotic terms and we expect the energy spectrum to be continuous.
\newline
\newline
I. Isotropic harmonic oscillator:
\newline
\newline
$Q_1^1:$
\newline
\begin{align*}
V(x,y)=&-\delta \hbar ^2 (x^2+y^2)+\frac{a}{x^2}+\hbar ^2 \big( \frac{\gamma}{P_5-1}+\frac{1}{y^2}(P_5-1)(\sqrt{2\alpha }+\alpha(2P_5-1)+\frac{\beta}{P_5})\nonumber\\
&+y^2(\frac{P_5'^2}{2 P_5}+\delta P_5)\frac{(2P_5-1)}{(P_5-1)^2}-\frac{P_5'}{P_5-1}-2\sqrt{2\alpha}P_5'\big)+\frac{3\hbar^2}{8y^2}.\nonumber\\
\end{align*}
$$Y_L=\{L_3^2,p_2^2\},$$
$$g_1(x,y)=2y( y W'+W+\frac{1}{3}\hbar ^2\delta y^3),\quad g_2(x,y)=-2x(3yW'+W+\frac{4}{3}\hbar ^2\delta y^3),$$
$$g_{3}(x,y)=x^2(4W'+2\hbar ^2\delta y^2)+\frac{2a}{x^2}y^2,$$
\begin{align*}
l(x,y)=& \hbar^2 x^2(\frac{1}{4}y W^{(4)}+W^{(3)})-x^2( 3y W'+W )W''-\hbar ^2 y(\frac{4}{3}\delta x^2y^2+\frac{3}{2})W''+(4(\frac{a}{x^2}-\hbar ^2\delta x^2)y^2-3 \hbar ^2)W'\nonumber\\
&+4y(\frac{a}{x^2}-\hbar ^2\delta x^2)W+\frac{4a}{3x^2}\hbar ^2\delta  y^4-2 \hbar ^2\delta x^2(\frac{2}{3} \hbar ^2\delta y^4- \hbar^2)-2 \hbar ^4\delta y^2.
\end{align*}
For
\begin{align*}
W(y)=&-\frac{\hbar ^2}{2y}\big(\frac{1}{P_{5}}(\frac{YP_5'}{P_{5}-1}-P_{5})^2-(1-\sqrt{2\alpha})^2(P_{5}-1)-2\beta \frac{P_{5}-1}{P_{5}}+\gamma Y \frac{P_{5}+1}{P_{5}-1}+2 \delta \frac{Y^2P_{5}}{(P_{5}-1)^2}\big)\\
&+\frac{\hbar ^2}{8y}-\frac{\delta \hbar^2}{3}y^3,
\end{align*}
where $P_5=P_5(Y); Y=y^2$.
\newline
\newline
$Q_1^2:$
\newline
\begin{align*}
V(x,y)=c_2( x^2+y^2)- \sqrt[4]{8 c_2^3 \hbar^2}y P_4(-\sqrt[4]{\frac{2c_2}{\hbar^2}}y) +\sqrt{\frac{c_2}{2}} \hbar( \epsilon  P_4'(-\sqrt[4]{\frac{2c_2}{\hbar^2}}y)+ P_4^2(-\sqrt[4]{\frac{2c_2}{\hbar^2}}y)); \quad \epsilon=\pm 1.\\
\end{align*}
$$Y_L=2p_1p_2^3,$$
$$g_1(x,y)=0,\;g_2(x,y)=3V-c_2(3x^2+y^2),\;g_{3}(x,y)=2c_2xy,$$
$$l(x,y)=-\frac{1}{4} \hbar^2 x V_{yyy}+3 x VV_y-c_2x(3x^2+ y^2)V_y.$$
\newline
\newline
II. Anisotropic harmonic oscillator:
\newline
\newline
$Q_{2}^1:$
\newline
\begin{align*}
V(x,y)=c_2(x^2+4y^2)+\frac{a}{x^2}-4 \sqrt[4]{2 c_2^3 \hbar^2}y P_4 +\sqrt{2c_2} \hbar( \epsilon  P_4'+ P_4^2); \quad \epsilon=\pm 1.\
\end{align*}
$$Y_L=\{L_3,p_1p_2^2\},$$
$$g_1(x,y)=-2 yW'-W+\frac{4}{3} c_2 y^3,\;g_2(x,y)=3xW'-4c_2 xy^2,\; g_{3}(x,y)=2c_2x^2y-2a\frac{y}{x^2},$$
\begin{align*}
l(x,y)=& -\frac{1}{8} \hbar^2 x^2 W^{(4)}+\frac{3}{2} x^2 W'W''-(2 c_2 x^2y^2-\frac{3 }{4}\hbar ^2)W''-2(2a \frac{y }{x^2}+2c_2 x^2 y)W'-2(\frac{a}{x^2}+c_2 x^2 )W\nonumber\\
&+\frac{8}{3}c_2y^3(c_2 x^2+\frac{a}{x^2})-2 c_2 \hbar^2y.\nonumber\\
\end{align*}
For
\begin{align*}
W(y)=\sqrt[4]{8c_2 \hbar ^6}\big(\frac{1}{8P_{4}}P_{4}'^2-\frac{1}{8}P_{4}^3-\frac{1}{2}Y P_{4}^2-\frac{1}{2}(Y^2-\alpha +\epsilon )P_{4}+\frac{1}{3}(\alpha -\epsilon )Y+\frac{\beta}{4P_{4}}\big)+\frac{4c_2}{3}y^3,
\end{align*}
where $P_4=P_4(Y);Y=-\sqrt[4]{\frac{8c_2}{\hbar ^2}} y.$
\newline
\newline
$Q_{2}^2:$
\newline
\begin{align*}
V(x,y)=&-\delta \hbar ^2 (4x^2+y^2) +\hbar ^2 \big( \frac{\gamma}{P_5-1}+\frac{1}{y^2}(P_5-1)(\sqrt{2\alpha }+\alpha(2P_5-1)+\frac{\beta}{P_5})\nonumber\\
&+y^2(\frac{P_5'^2}{2 P_5}+\delta P_5)\frac{(2P_5-1)}{(P_5-1)^2}-\frac{P_5'}{P_5-1}-2\sqrt{2\alpha}P_5'\big)+\frac{3\hbar^2}{8y^2}.
\end{align*}
$$Y_L=\{L_3,p_2^3\},$$
$$g_1(x,y)=0,\;g_2(x,y)=-3yW'-W-\frac{4}{3}\hbar ^2 \delta y^3,\;g_{3}(x,y)=4xW'+4\hbar ^2 \delta xy^2,$$
$$l(x,y)=\frac{1}{4} \hbar^2 x (y W^{(4)}+4 W^{(3)})-3 x y W'^2-xWW'-\frac{4}{3}\hbar ^2 \delta x y^3 W''+2 \hbar^4 \delta x.$$
\newline
For
\begin{align*}
W(y)=&-\frac{\hbar ^2}{2y}\big(\frac{1}{P_{5}}(\frac{YP_5'}{P_{5}-1}-P_{5})^2-(1-\sqrt{2\alpha})^2(P_{5}-1)-2\beta \frac{P_{5}-1}{P_{5}}+\gamma Y \frac{P_{5}+1}{P_{5}-1}+2 \delta \frac{Y^2P_{5}}{(P_{5}-1)^2}\big)\\
&+\frac{\hbar ^2}{8y}-\frac{4\delta \hbar^2}{3}y^3,
\end{align*}
where $P_5=P_5(Y); Y=y^2$.
\newline
\newline
III. Potentials with no confining (harmonic oscillator) term:
\newline
\newline
$Q_{3}^1:$
\newline 
\begin{align*}
V(x,y)=&\frac{a}{x^2}+\frac{\hbar ^2}{2}(\sqrt{\alpha } P_3'+\frac{3}{4} \alpha  (P_3)^2+\frac{\delta }{4P_3^2}+\frac{\beta P_3}{2y}+\frac{\gamma}{2y P_3}-\frac{ P_3'}{2y P_3}+\frac{P_3'^2}{4 P_3^2}).\nonumber\\
\end{align*}
$$Y_L=\{L_3^2,p_2^2\},$$
$$g_1(x,y)=2 y^2W'+2 y W,\;g_2(x,y)=-6xyW'-2xW,\; g_{3}(x,y)=4x^2W'+2a\frac{y^2}{x^2},$$
$$l(x,y)=\hbar^2 x^2(\frac{1}{4}y W^{(4)}+W^{(3)})-x^2( 3y W'+W )W''-\frac{3}{2} \hbar^2 yW''+(4\frac{a}{x^2}y^2-3 \hbar^2)W'+4\frac{a}{x^2}yW.$$
\newline
For
\begin{align*}
W(y)=&-\frac{\hbar ^2}{2y}\big(\frac{1}{4}(y\frac{P_3'}{P_3}-1)^2-\frac{1}{16} \alpha y^2 P_3^2-\frac{1}{8}(\beta+2\sqrt{\alpha})y P_3+\frac{\gamma}{8P_3} y+\frac{\delta}{16 P_3^2} y^2\big)+\frac{\hbar ^2}{8y}.
\end{align*}
\newline
\newline
$Q_{3}^2:$
\newline
\begin{align*}
V(x,y)=&\frac{a}{x^2}+\frac{b^2 \hbar ^2}{2}( \epsilon  P_2'+ P_2^2); \quad \epsilon=\pm 1.\nonumber\\
\end{align*}
$$Y_L=\{L_3,p_1p_2^2\},$$
$$g_1(x,y)=-2 yW'-W-\frac{b^3 \hbar ^2}{8}y^2,\;g_2(x,y)=3xW'+\frac{b^3 \hbar ^2}{4}xy, \;g_{3}(x,y)=-\frac{b^3 \hbar ^2}{8}x^2-2a\frac{y}{x^2},$$
\begin{align*}
l(x,y)=-\frac{1}{8} \hbar^2 x^2 W^{(4)}+\frac{3}{2} x^2 W'W''+(\frac{b^3 \hbar ^2}{8} x^2 y+\frac{3 }{4}\hbar ^2)W''-4a \frac{y }{x^2}W'-2\frac{a}{x^2}W-ab^3 \hbar ^2 \frac{ y^2}{4x^2}.\nonumber\\
\end{align*}
For
\begin{align*}
W(y)=\frac{-b\hbar ^2}{2}\big((P_2')^2-(P_2^2+\frac{b}{2}y)^2-2(\alpha+\frac{\epsilon}{2})P_2\big)-\frac{b^3}{8}\hbar ^2 y^2,\nonumber\\
\end{align*}
where $P_2=P_2(by).$
\newline
\newline
$Q_{3}^3:$
\newline
$$V(x,y)=\frac{a}{x^2}+\hbar ^2 b^2 P_1.$$
$$Y_L=\{L_3^2,p_2^2\},$$
$$g_1(x,y)=-2 yW'-W,\;g_2(x,y)=3xW',\; g_{3}(x,y)=-2a\frac{y}{x^2},$$
$$l(x,y)= -\frac{1}{8} \hbar^2 x^2 W^{(4)}+\frac{3}{2} x^2 W'W''+\frac{3 }{4}\hbar ^2W''-4 a \frac{y }{x^2}W'-2\frac{a}{x^2}W$$
\newline
For
$$W(y)=-b\hbar ^2(\frac{1}{2}(P_1')^2-2P_1^3-b y P_1),$$
where $P_1=P_1(by).$
\newline
\newline
$Q_{3}^4:$
\newline
$$V(x,y)=\frac{a}{x^2}+\hbar ^2 \wp.$$
$$Y_L=\{L_3^2,p_2^2\},$$
$$g_1(x,y)=-2 yW'-W,\;g_2(x,y)=3xW',\; g_{3}(x,y)=-2a \frac{y}{x^2},$$
$$l(x,y)= -\frac{1}{8} \hbar^2 x^2 W^{(4)}+\frac{3}{2} x^2 W'W''+\frac{3 }{4}\hbar ^2W''-4 a \frac{y }{x^2}W'-2\frac{a}{x^2}W$$
\newline
For
$$W(y)=\hbar ^2 \int{u}dy ,\quad u=\wp(y).$$
\newline
\newline
$Q_{3}^5:$
\newline
\begin{align*}
V(x,y)=&c_1x+\frac{\hbar ^2}{2}(\sqrt{\alpha } P_3'(y)+\frac{3}{4} \alpha  (P_3(y))^2+\frac{\delta }{4P_3^2(y)}+\frac{\beta P_3(y)}{2y}+\frac{\gamma}{2y P_3(y)}-\frac{ P_3'(y)}{2y P_3(y)}+\frac{P_3'^2(y)}{4 P_3^2(y)}).\nonumber\\
\end{align*}
$$Y_L=\{L_3,p_2^3\},$$
$$g_1(x,y)=0,\;g_2(x,y)=-3yW'-W,\;g_{3}(x,y)=4xW'-\frac{1}{2}c_1y^2,$$
$$l(x,y)=\frac{1}{4} \hbar^2 x (y W^{(4)}+4 W^{(3)})-3 x y W'^2-xWW'-c_1 y^2 W'-c_1 y W.$$
For
\begin{align*}
W(y)=&-\frac{\hbar ^2}{2y}\big(\frac{1}{4}(y\frac{P_3'}{P_3}-1)^2-\frac{1}{16} \alpha y^2 P_3^2-\frac{1}{8}(\beta+2\sqrt{\alpha})y P_3+\frac{\gamma}{8P_3} y+\frac{\delta}{16 P_3^2} y^2\big)+\frac{\hbar ^2}{8y}.
\end{align*}
\newline
\newline
$Q_{3}^6:$
\newline
$$V(x,y)=c_1x+\frac{b^2 \hbar ^2}{2}( \epsilon  P_2'+ P_2^2); \quad \epsilon=\pm 1.$$
$$Y_L=2p_1p_2^3,$$
$$g_1(x,y)=0,\;g_2(x,y)=3W'+\frac{b^3\hbar ^2}{4}y,\;g_{3}(x,y)=c_1y-\frac{b^3\hbar ^2}{4}x,$$
$$l(x,y)=-\frac{1}{4} \hbar^2 x W^{(4)}+3 x W'W''+\frac{b^3\hbar ^2}{4} x y W''+2c_1 y W'+c_1 W+\frac{b^3\hbar ^2}{8} c_1 y^2.$$
For
\begin{align*}
W(y)=\frac{-b\hbar ^2}{2}\big((P_2')^2-(P_2^2+\frac{b}{2}y)^2-2(\alpha+\frac{\epsilon}{2})P_2\big)-\frac{b^3}{8}\hbar ^2 y^2,\nonumber\\
\end{align*}
where $P_2=P_2(by).$
\newline
\newline
$Q_{3}^7:$
\newline
$$V(x,y)=c_1x+\hbar ^2 b^2 P_1.$$
$$Y_L=2p_1p_2^3,$$
$$g_1(x,y)=0,\;g_2(x,y)=3W',\;g_{3}(x,y)=c_1y,\; l(x,y)=-\frac{1}{4} \hbar^2 x W^{(4)}+3 x W'W''+2c_1 y W'+c_1 W.$$
\newline
For
$$W(y)=-b\hbar ^2(\frac{1}{2}(P_1')^2-2P_1^3-b y P_1),$$
where $P_1=P_1(by).$
\newline
\newline
$Q_{3}^8:$
\newline
$$V(x,y)=c_1x +\hbar ^2 \wp.$$
$$Y_L=2p_1p_2^3,$$
$$g_1(x,y)=0,\;g_2(x,y)=3W',\;g_{3}(x,y)=c_1y,\; l(x,y)=-\frac{1}{4} \hbar^2 x W^{(4)}+3 x W'W''+2c_1 y W'+c_1 W.$$
\newline
\newline
For
$$W(y)=\hbar ^2 \int{u}dy ,\quad u=\wp(y).$$
\newline
\newline
The potentials $Q_1^2, Q_3^6$ and $Q_3^7$ are in the list of quantum potentials obtained by Gravel \cite[($Q_{18},Q_{19}, Q_{21}$)]{Gravel:G3}. Among the integrals of motion we have $\{L_3^2,p_2^2\}$ and $\{L_3,p_2^3\}$. These can not be obtained by commuting a third and a second order integral. Marquette in \cite{Marquette:P5} obtained a potential in terms of fifth Painl\'eve transcendent for a system admitting fourth order ladder operators which allowed a characterisation of the spectrum and wave functions in a recursive way from the zero modes and
build integrals for families of 2D models.
\subsection{FUTURE OUTLOOK}
Part II of this article will follow shortly and will be devoted to a complete analysis of the nonexotic potentials. They are obtained when the linear compatibility conditions (\ref{*(1&2)}) and (\ref{**(1&2)}) are not satisfied identically. They must then be solved as ODEs.\\
We are also currently studying whether some or possibly all exotic potentials can be generated from one-dimensional Hamiltonians using algebras of differential operators depending on one variable only.

\section*{ACKNOWLEDGEMENTS}
The research of P.W. was partially supported by an NSERC discovery grant. M.S. thanks the University of Montreal for a "bourse d'admission" and a "bourse de fin d'\'etudes doctorales". I.M. was supported by the Australian Research Council through Discovery Early Career Researcher Award DE130101067. Also the authors thank R.Conte for very helpful discussions.
\maketitle

\newcommand{\etalchar}[1]{$^{#1}$}


\end{document}